\documentclass[pra,aps,twocolumn,superscriptaddress]{revtex4}

\usepackage{graphicx,amsmath}
\usepackage{txfonts}
\usepackage{color}
\usepackage{hyperref}

\newcommand{\rp}[1]{(\ref{#1})}

\newcommand{\abs}[1]{\left|{#1}\right|}

\newcommand{\av}[1]{\left\langle #1 \right\rangle}

\newcommand{\wt}[0]{\widetilde}

\newcommand{\al}[1]{^{(#1)}}
\newcommand{\da}{^\dagger}

\newcommand{\pt}[1]{\left( #1 \right)}
\newcommand{\pq}[1]{\left[ #1 \right]}
\newcommand{\pg}[1]{\left\{ #1 \right\}}

\newcommand{\lpq}[1]{\left[ #1 \right.}
\newcommand{\lpg}[1]{\left\{ #1 \right.}

\newcommand{\rpq}[1]{\left. #1 \right]}
\newcommand{\rpg}[1]{\left. #1 \right\}}
\newcommand{\ee}{{\rm e}}
\newcommand{\ii}{{\rm i}}
\newcommand{\dd}{{\rm d}}

\newcommand{\id}{\openone}

\newcommand{\nn}{{\nonumber}}

\newcommand{\mat}[2]{
                      \begin{array}{#1}
                       #2
                       \end{array}  }

\newcommand{\mm}[2]{ \pt{
                      \begin{array}{cc}
                       #1 \\
                       #2
                     \end{array}  }  }

\newcommand{\va}{{\bf a}}

\newcommand{\CC}{{\cal C}}

\newcommand{\MM}{{\cal M}}

\newcommand{\SSS}{{\cal S}}

\newcommand{\fb}{{\rm fb}}
\newcommand{\un}{{\rm un}}
\newcommand{\om}{{\rm om}}
\newcommand{\eff}{{\rm eff}}
\newcommand{\out}{{\rm out}}
\newcommand{\inn}{{\rm in}}

\begin{document}

\title{
Cavity optomechanics with feedback--controlled in--loop light
}

\author{Stefano Zippilli}
\affiliation{School of Science and Technology, Physics Division, University of Camerino, I-62032 Camerino (MC), Italy}
\affiliation{INFN, Sezione di Perugia, I-06123 Perugia, Italy}
\author{Nenad Kralj}
\affiliation{School of Science and Technology, Physics Division, University of Camerino, I-62032 Camerino (MC), Italy}
\author{Massimiliano Rossi}
\affiliation{Niels Bohr Institute, University of Copenhagen, 2100 Copenhagen, Denmark}
\affiliation{Center for Hybrid Quantum Networks (Hy-Q), Niels Bohr Institute, University of Copenhagen, 2100 Copenhagen, Denmark}
\author{Giovanni Di Giuseppe}
\affiliation{School of Science and Technology, Physics Division, University of Camerino, I-62032 Camerino (MC), Italy}
\affiliation{INFN, Sezione di Perugia, I-06123 Perugia, Italy}
\author{David Vitali}
\affiliation{School of Science and Technology, Physics Division, University of Camerino, I-62032 Camerino (MC), Italy}
\affiliation{INFN, Sezione di Perugia, I-06123 Perugia, Italy}
\affiliation{CNR-INO, Largo Enrico Fermi 6, I-50125 Firenze, Italy}

\date{\today}

\begin{abstract}
It has recently been shown [Rossi {\it et al.}, Phys. Rev. Lett. {\bf 119}, 123603 (2017); {\it ibid.} {\bf 120}, 073601 (2018)] that feedback--controlled in--loop light can be used to enhance the efficiency of optomechanical systems. We analyse the theoretical ground at the basis of this approach and explore its potentialities and limitations.
We discuss the validity of the model, analyse the properties of in-loop cavities and we show how they can be used
to observe coherent optomechanical oscillations also with a weakly coupled system, improve the sideband cooling performance, and increase ponderomotive squeezing.
\end{abstract}
\maketitle


\section{Introduction}

In cavity optomechanics~\cite{Bowen,Aspelmeyer} the radiation pressure interaction between a resonant mode of an optical cavity and a vibrational mode of a mechanical resonator is exploited for sensitive measurements~\cite{Metcalfe}, storage and transduction of light signals~\cite{Fiore2011}, and as a test bed for the investigation of nonlinear dynamics~\cite{Marquardt2006}. When operated at the quantum level, cavity optomechanical devices allow for the engineering of quantum mechanical dynamics that may find applications in quantum information processing~\cite{Stannigel}, and in the study of macroscopic quantum effects~\cite{Chen2013}.
In this context measurement--based feedback~\cite{Serafini2012,Zhang2017} has been discussed as a useful tool for the engineering  of quantum states of the mechanical resonator and for enhanced sensing. Specific implementations~\cite{Wilson2015,Sudhir,Schafermeier,Rossi2018} rely on the use of light fields as part of the detector and of the actuator for the feedback loop that operates directly on the mechanical element.

Here we explore a different approach. Specifically, we use a feedback loop to engineer the light fluctuations of the laser field which drives the system (see also a related proposal with electromechanical systems~\cite{Zhang2009}). 
Feedback--controlled in--loop fields have been studied as a means to
reduce light fluctuations (so called light squashing)~\cite{Shapiro87}.
While squashing can not be extracted out of the feedback loop, so
that it is different from quantum squeezing~\cite{Shapiro87}, useful
applications of in-loop light have been discussed. The central observation is that in-loop light can be useful when employed to drive and manipulate the dynamics of an additional system. By this means, the out-of-loop response of the additional system can be improved.
This was first suggested theoretically in
\cite{Yamamoto86},
and discussed also in
\cite{Shapiro87},
where it is shown that QND detection can be used to extract squeezed light from an in-loop squashed field. In these works an additional Kerr-medium is driven by the in-loop field and the out-of-loop response of the medium exhibits quantum properties. A second notable example is presented in Refs.
\cite{Wiseman98,Wiseman99},
where it is predicted that an atom responds to in-loop light in a way similar to what is expected for squeezed light.
The recent works reported in Refs.~\cite{Rossi,Kralj,Rossi2}
demonstrate the feasibility of similar approaches with an optomechanical system, showing that
feedback--controlled light
can be employed to tune at will the response of a mechanical system.
More specifically, these works show that
in-loop optical fields can be properly tailored to
enhance the efficiency of
laser cooling even beyond the back--action limit, and
to promote a naturally weakly coupled system to the strong coupling regime by effectively reducing the cavity linewidth.

In this work we discuss in detail the theoretical model used to describe these systems, and show that feedback--controlled light may play a significant role as a novel efficient tool for manipulating cavity--optomechanical devices.
In particular, we review the basic ideas of squashed and anti-squashed light,
and demonstrate how in--loop light can exhibit reduced fluctuations at specific frequencies which
can be exploited to tailor the light scattering rates of a mechanical resonator.
We further show that the dynamics of an in-loop cavity can be modelled by a standard cavity
with an effectively reduced or enhanced cavity decay rate.
Then we discuss how these facts allow
to improve resolved sideband cooling and
enter the strong coupling regime even in a weakly coupled system.
Finally, we describe how the feedback that operates by measuring the light leaking through a cavity output may be properly engineered to enhance the ponderomotive squeezing of the light leaking through another cavity output.

The article is structured as follows. In Sec.~\ref{SecLight} we introduce the feedback model that operates on a laser field. In Sec.~\ref{SecCav} we analyse the feedback when an optical cavity is added within the loop. Then, in Sec.~\ref{SecOM} we include also a mechanical resonator, and we study in detail the dynamics of the optomechanical system, including optomechanical oscillations, cooling and ponderomotive squeezing. Finally, in Sec.~\ref{conclusion} we draw our conclusions and discuss some possible outlooks.

\section{Feedback--controlled Light}\label{SecLight}

\begin{figure}[t!]
\includegraphics[width=8.5cm]{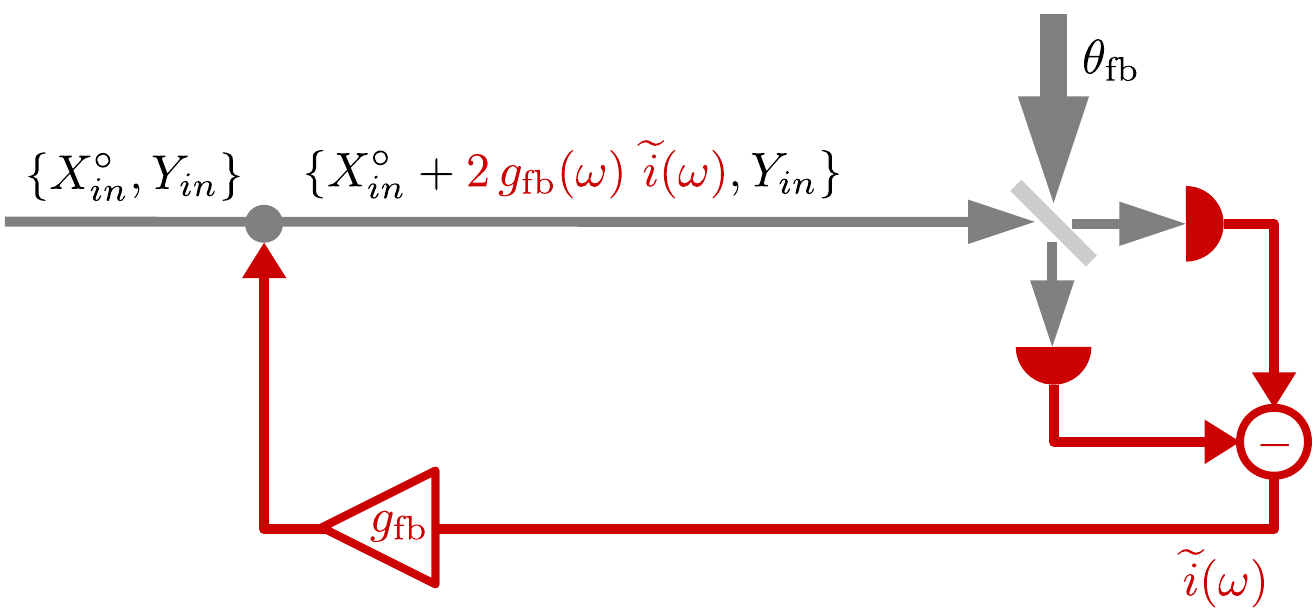}
\caption{The feedback loop: a field quadrature at phase $\theta_\fb$ is detected, and the corresponding photocurrent is used to modulate the amplitude $X_\inn$ of the field itself, while the field phase $Y_\inn$ remains unaffected.
}\label{schema0}
\end{figure}

In this section we introduce the basic elements of the feedback model.
In particular we study the squashing and anti--squashing of light that is observed in the simple situation in which
a laser field is detected (either by direct photodetection or homodyne detection) and the recorded signal is used to modulate the field amplitude~\cite{Shapiro87} as in Fig.~\ref{schema0}.

We consider a continuous wave field~\cite{ZippilliNJP15} described by the
electric field $E(t)\propto \ee^{-\ii\,\omega_L\,t}\ A_{\rm in}(t)+h.c.$ with
annihilation operator $A_{\rm in}(t)$ which we
decompose in terms of the coherent amplitude $\alpha_{\rm in}(t)$ and the operator for the quantum fluctuations $a_{\rm in}(t)$ such that
\begin{eqnarray}
A_{\rm in}(t)=\alpha_{\rm in}(t)+a_{\rm in}(t)\ ,
\end{eqnarray}
with $\av{a_{\rm in}}=0$.
Similarly we decompose the detected photocurrent
\begin{eqnarray}\label{I}
I(t)=\bar I(t)+i(t)
\end{eqnarray}
in terms of amplitude $\bar I(t)$ and fluctuations $i(t)$, with $\av{i(t)}=0$, the specific form of which is reported below in Eq.~\rp{i}.
The photocurrent is utilized to modulate 
the input field according to the relation
\begin{eqnarray}
A_{\rm in}(t)&=&A_{\rm in}^\circ(t)+F_{\rm fb}(t),
\end{eqnarray}
where the symbol $^\circ$ indicates quantities with no feedback and the term $F_{\rm fb}$ describes the effect of feedback, explicitly given by
\begin{eqnarray}\label{Ffb}
F_{\rm fb}(t)&=&\frac{1}{\sqrt{2\,\pi}}\int_{t_0}^t\dd t'\ g_{\rm fb}(t-t')\ I(t'),
\end{eqnarray}
with $g_{\rm fb}(t-t')$ a causal filter function, meaning that it is zero for $t'>t$, hence the upper limit of integration can be extended to infinity.
We also note that, in general, the filter function includes a finite delay $\tau_{\rm fb}$ so that the feedback
does not act instantaneously on the input field, and $g_{\rm fb}(t)$ is proportional to the step function $\theta(t-\tau_{\rm fb})$.
Finally we decompose also this expression in terms of amplitude and fluctuations according to $F_{\rm fb}(t)=\bar F_{\rm fb}(t)+\Phi(t)$ with
\begin{eqnarray}
\bar F_{\rm fb}(t)&=&\frac{1}{\sqrt{2\,\pi}}\int_{t_0}^\infty\dd t'\ g_{\rm fb}(t-t')\ \bar I(t')\ ,
\nn\\
\Phi(t)&=&\frac{1}{\sqrt{2\,\pi}}\int_{t_0}^\infty\dd t'\ g_{\rm fb}(t-t')\ i(t')\ ,
\label{Phi}
\end{eqnarray}
such that the field amplitude and fluctuations are respectively given by
\begin{eqnarray}\label{alphainain}
\alpha_{\rm in}(t)&=&\alpha_{\rm in}^\circ+\bar F_{\rm fb}(t),
\nn\\
a_{\rm in}(t)&=&a_{\rm in}^\circ(t)+\Phi(t)\ .
\end{eqnarray}

\subsection{In-loop photocurrent}

We consider the situation in which the feedback response function $g_\fb(t)$ realizes a high--pass filter, which cuts the low frequency components of the photocurrent that correspond to the coherent part of the light signal as in Ref.~\cite{Rossi,Kralj,Rossi2}. In this case the average photocurrent remains constant and we are interested only in the dynamics of the fluctuations $i(t)$.

The effects of detection inefficiencies and electronic noise can be modelled in terms of a perfect detector preceded by a beam splitter with finite transmissivity $\sqrt{\eta}$, such that the fluctuations of the photocurrent can be expressed as
\begin{eqnarray}\label{i}
i(t)&=&\sqrt{\eta}\ X_{\rm in}\al{\theta_\fb }(t)+\sqrt{1-\eta}\ X_v(t)\ ,
\end{eqnarray}
where we have introduced the detected field quadrature at phase $\theta_\fb $
\begin{eqnarray}\label{Xintheta}
X_{\rm in}\al{\theta_\fb }(t)&=&\ee^{-\ii\theta_\fb }\,a_{\rm in}(t)+\ee^{\ii\theta_\fb }\,a_{\rm in}(t)\da\ ,
\end{eqnarray}
and where $X_v(t)$ is the noise operator which accounts for the additional noise due to inefficient detection. Here we assume that the photocurrent is properly normalized so that the photocurrent power spectrum of a coherent field is set to one (which, hence, corresponds to the level electronic plus shot noise). In particular this implies that $X_v(t)$ fulfils the relation $\av{X_v(t)\ X_v(t')}=\delta(t-t')$.
Moreover, the effective detection efficiency $\eta$ is related to the real detection efficiency $\eta_d$ (which comprises both the detector quantum efficiency $\eta_Q$ and the optical path efficiency $\eta_O$, i.e. $\eta_d=\eta_Q\,\eta_O$) by the relation $\eta=\eta_d/(1+S_e/S_{sn})$, where $S_e/S_{sn}$ is the ratio between electronic and shot noise~\cite{electronicNoise}.
This model approximates, retaining only linear terms in the field fluctuations, both homodyne detection (in the limit of large amplitude of the local oscillator, and with the phase difference between signal and local oscillator equal to $\theta_\fb$), and direct photodetection for the special case $\theta_\fb=0$ (that is valid in the limit of large amplitude of the signal itself).

Including the field quadrature without feedback $ X_{\rm in}^{\circ\,\pt{\theta_\fb}}(t)$, and the corresponding photocurrent $i^\circ(t)=\sqrt{\eta}\ X_{\rm in}^{\circ\,\pt{\theta_\fb }}(t)+\sqrt{1-\eta}\ X_v(t)$, we find that
Eq.~\rp{i} can be rewritten as
\begin{eqnarray}
i(t)&=&i^\circ(t)+\sqrt{\eta}\pq{\ee^{-\ii\,\theta_\fb }\ \Phi(t)+\ee^{\ii\,\theta_\fb }\ \Phi(t)^* } \ .
\end{eqnarray}

\subsubsection{Power spectrum of the in-loop photocurrent}\label{squashing}


\begin{figure}[h!]
\includegraphics[width=7.cm]{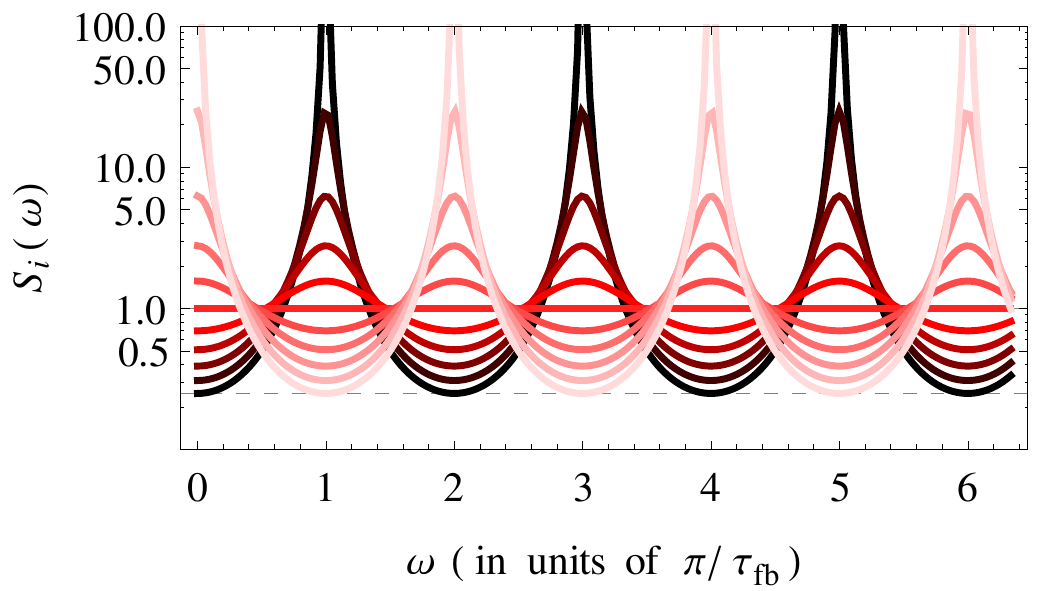}
\caption{Photocurrent power spectrum normalized such that the power spectrum for a coherent field is equal to one, and evaluated for the filter function defined in Eq.~\rp{gfbflat} (with $\phi_\fb=0$). Lines from dark to light red
correspond to values of $\bar g_\fb$ which range from $\bar g_\fb=-1/(2\sqrt{\eta}\,\cos\theta_\fb)$ to $\bar g_\fb=1/(2\sqrt{\eta}\,\cos\theta_\fb)$, and the horizontal 
red solid
line corresponds to $\bar g_\fb=0$. The maxima of both the lightest and darkest
curves diverge. The minima are indicated by the horizontal dashed line at the value $1/4$.
}\label{lambda0}
\end{figure}

Let us now study the stationary properties of the in-loop photocurrent $i(t)$.
We assume that the initial time $t_0$ introduced in Eq.~\rp{Ffb} is in the far past so that we approximate the expression for the feedback term with $t_0\to-\infty$, i.e $\Phi(t)=\frac{1}{\sqrt{2\,\pi}}\int_{-\infty}^\infty\dd t'\ g_{\rm fb}(t-t')\ i(t')$.
Defining the Fourier transform as
$\wt x(\omega)=\frac{1}{\sqrt{2\pi}}\int\dd t\ \ee^{\ii\omega t}\,x(t)$,
the expression for the photocurrent in Fourier space becomes
$\wt i(\omega)=
\wt i^\circ(\omega)+\sqrt{\eta}\pq{\wt g_{\rm fb}(\omega)\,\ee^{-\ii\,\theta_\fb}+\wt g_{\rm fb}(-\omega)^*\,\ee^{\ii\,\theta_\fb}}\,\wt{i}(\omega)$.
When the feedback modulates the amplitude of the input field then $g_\fb(t)$ is real, while if the feedback modulates the phase then $g_\fb(t)$ is imaginary. In general it is possible to decompose the feedback function as 
$g_\fb(t)=g_{\fb,A}(t)+\ii\,g_{\fb,P}(t)$, where $g_{\fb,A}(t)$ and $g_{\fb,P}(t)$ are real and account for the effect of the feedback on, respectively, the amplitude and phase of the driving field. Thereby, in Fourier space we find $\wt g_{\fb,A}(\omega)=\wt g_{\fb,A}(-\omega)^*$ and $\wt g_{\fb,P}(\omega)=\wt g_{\fb,P}(-\omega)^*$, so that
$\wt i(\omega)=
\wt i^\circ(\omega)+2\,\sqrt{\eta}\pq{
\cos\pt{\theta_\fb}\,\wt g_{\rm fb,A}(\omega)
+\sin\pt{\theta_\fb}\,\wt g_{\rm fb,P}(\omega)
}\,\wt{i}(\omega)$.
Here we focus on the situation in which the feedback modulates only the amplitude of the driving field. However we note that the study that we report hereafter can be easily extended to the general case using the previous expression for the photocurrent. In particular, here, we use $g_{\fb,A}(t)=g_\fb(t)$ so that the photocurrent reduces to
\begin{eqnarray}\label{ifb00}
\wt i(\omega)&=&\wt i^\circ(\omega)+2\sqrt{\eta}\cos\pt{\theta_\fb }\ \wt g_{\rm fb}(\omega)\,\wt{i}(\omega)\ .
\end{eqnarray}
We further note that this expression describes also the case in which the feedback functions for the amplitude and the phase are proportional to each other $g_{\fb,A}(t)\propto g_{\fb,B}(t)$. In this case, in fact, the complex feedback function $g_\fb(t)$ exhibits a constant phase, i.e. $g_\fb(t)=\abs{g_\fb(t)}\,\ee^{\ii\phi_{g_\fb}}$, so that the photocurrent is similar to Eq.~\rp{ifb00} but with the substitution $\theta_\fb \to \theta_\fb-\phi_{g_\fb}$.

The solution of Eq.~\rp{ifb00} can be expressed in terms of the squashing factor
\begin{eqnarray}\label{lambdafree}
\lambda(\omega)=\frac{1}{1-2\sqrt{\eta}\ \wt g_{\rm fb}(\omega)\ \cos\pt{\theta_\fb }}
\end{eqnarray}
which fulfils the relation $\lambda(\omega)=\lambda^*(-\omega)$, as
\begin{eqnarray}\label{ifree}
\wt i(\omega)
&=&\wt i^\circ(\omega)\ \lambda(\omega)\ .
\end{eqnarray}
Finally the power spectral density of the photocurrent $S_i(\omega)$ is defined by the relation $\av{\wt i(\omega)\ \wt i(\omega')}=S_i(\omega)\,\delta(\omega+\omega')$. Thus, when the input field is a coherent laser field it is given by
\begin{eqnarray}
S_i(\omega)&=&\abs{\lambda(\omega)}^2\ ,
\end{eqnarray}
where we have used the fact that in this case, according to our normalization, the power spectral density in the absence of feedback is equal to $S^\circ_i(\omega)=1$, which corresponds to  the level of electronic plus shot noise as discussed above. We note that when one measures the phase quadrature, $\theta_\fb =\frac{\pi}{2}$, the loop, that acts on the orthogonal, amplitude quadrature, is not closed and the corresponding power spectrum is equal to one.


In order to gain insight into the behaviour of the feedback photocurrent,
we consider here a specific form of the feedback filter function $\wt g_\fb(\omega)$.
Apart form very low frequencies not interesting for our purpose, where as specified above the filter $\wt g_\fb(\omega)$ is zero (high--pass filter), we assume that in the relevant band of frequencies the filter function is constant, with a linear change in phase due to a finite feedback delay time $\tau_{\rm fb}$, that is
\begin{eqnarray}\label{gfbflat}
\wt g_{\rm fb}(\omega)&=&\bar g_{\rm fb}\ \ee^{\ii\,\omega\,\tau_{\rm fb}+\phi_\fb\,{\omega}/{\abs{\omega}}}\ ,
\end{eqnarray}
where we have also included a phase offset $\phi_\fb$, and where the term ${\omega}/{\abs{\omega}}$ is needed in order to satisfy the relation $\wt g_\fb(\omega)=\wt g_\fb(-\omega)^*$.
Thereby we find
\begin{eqnarray}
S_i(\omega)&=&
\lpq{1-4\,\sqrt{\eta}\,\cos\pt{\theta_\fb }\ \bar g_{\rm fb}\cos\pt{\omega\tau_\fb
 +\phi_\fb
}
}\nn\\&&\rpq{
+4\,\eta\ \bar g_\fb^2\ \cos\pt{\theta_\fb }^2}^{-1}\ .
\end{eqnarray}
The power spectrum $S_i(\omega)$ is reduced below one, corresponding to light squashing, for negative feedback
$\cos(\omega\,\tau_\fb+\phi_\fb)<2\sqrt{\eta}\abs{\bar g_\fb\ \cos(\theta_\fb)}$, while it is enhanced (anti-squashed light) for positive feedback, $\cos(\omega\,\tau_\fb+\phi_\fb)>2\sqrt{\eta}\abs{\bar g_\fb\ \cos(\theta_\fb)}$ (see Fig.~\ref{lambda0}). In particular
the spectrum exhibits peaks which
diverge when $\bar g_\fb\,\cos\pt{\theta_\fb }\to {1}/{(2\sqrt{\eta})}$. It follows that the feedback is unstable for $\abs{\bar g_\fb\,\cos\pt{\theta_\fb }}\geq{1}/{(2\sqrt{\eta})}$.
Within the regime of stability $\abs{\bar g_\fb\,\cos\pt{\theta_\fb }}<{1}/{(2\sqrt{\eta})}$,
the maxima and minima of the power spectrum are found at frequencies
\begin{eqnarray}\label{omegamaxmin}
\omega_n=\pm\frac{n\,\pi
 -\phi_\fb
}{\tau_\fb}\ ,
\end{eqnarray}
with integer $n\geq0$,
and the corresponding values are
\begin{eqnarray}
S_i\pt{\omega_n}=\lambda(\omega_n)^2=\frac{1}{\pq{1-(-1)^n\ 2\,\sqrt{\eta}\ \bar g_\fb\ \cos\pt{\theta_\fb }}^2}\ .
\end{eqnarray}
Thus, assuming, for example
$\bar g_\fb\,\cos(\theta_\fb )>0$, 
the maxima (minima) are found for even (odd) n.
In particular, the minimum value is achieved at these frequencies in the limit $\bar g_\fb\to \frac{1}{2\,\sqrt{\eta}\,\cos(\theta_\fb )}$, and it is given by ${\rm min}\pq{S_i\pt{\omega_{\omega_{2n+1}}}}=1/4$.
We finally remark that if the function $\wt g_\fb(\omega)$ corresponds to a bandpass filter, and
the delay time is sufficiently short, one can set the feedback phases so that no maxima fall within the feedback bandwidth. In this case the amplitude of $\wt g_\fb(\omega)$ can be increased indefinitely and the minimum can approach the value zero (in the limit of infinite negative feedback).

\subsection{The in-loop field}\label{comm}

Let us now study the properties of the in-loop field.
%
It is important to note that the in-loop field is not a free field and its operators do not fulfil the standard bosonic commutation relations~\cite{Shapiro87}. This can be shown as follows. The feedback relation in Eq.~\rp{alphainain} can be expressed in the frequency domain as
\begin{eqnarray}\label{ain}
\wt a_{\rm in}(\omega)=\wt a_{\rm in}^\circ(\omega)+\wt g_\fb(\omega)\ \wt i(\omega)
\end{eqnarray}
where
the field $\wt a_{\rm in}^\circ(\omega)$ is free and does fulfil the standard bosonic commutation relation $\pq{\wt a_{\rm in}^\circ(\omega),\wt a_{\rm in}^\circ{}\da(\omega)}=\delta(\omega+\omega')$ [note that in this work we use the notation according to which, given an operator in Fourier space $\wt o(\omega)$,   $\pg{\wt o(\omega)}\da\equiv \wt o\da(-\omega)$]. For the in-loop field, instead,
using Eq.~\rp{ifree} we find
\begin{eqnarray}
\pq{\wt a_{\rm in}(\omega),\wt a_{\rm in}\da(\omega')}
&=&\delta(\omega+\omega')
\\&&\hspace{-0.5cm}\times
\pg{
1+2\,\sqrt{\eta}\,{\rm Re}\pq{\wt g_\fb(\omega)\,\lambda(\omega)\,\ee^{-\ii\,\theta_\fb }}
}\nn
\end{eqnarray}
 and
\begin{eqnarray}
\pq{\wt a_{\rm in}(\omega),\wt a_{\rm in}(\omega')}
&=&-\delta(\omega+\omega')
\\&&\hspace{-0.cm}\times
\ 2\ \ii\, \sqrt{\eta}\, {\rm Im}\pq{\wt g_\fb(\omega)\,\lambda(\omega)}\,\ee^{\ii\,\theta_\fb }\ .
\nn
\end{eqnarray}
We also highlight that, when analysed in the time domain, the standard bosonic commutation relations
are recovered for two operators at a time difference smaller than the feedback delay time. In this case, in fact,
the field behaves as a free field~\cite{Shapiro87}. This can be seen by calculating the inverse Fourier transform of the previous expressions, such as 
$\pq{a_{\rm in}(t), a_{\rm in}\da(t')}=
\frac{1}{2\,\pi}\int_{-\infty}^{\infty}\dd\omega\ 
\int_{-\infty}^{\infty}\dd\omega'\ 
\ee^{\ii\pt{\omega\, t+\omega'\, t'}}
\pq{\wt a_{\rm in}(\omega),\wt a_{\rm in}\da(\omega')}
$,
exploiting the analytic properties of the causal filter function $\wt g_\fb(\omega)$ (a causal function is analytic in the upper half complex plane).
Specifically, this can be done by expanding the term $\wt g_\fb(\omega)\,\lambda(\omega)$ in powers of $\wt g_\fb(\omega)$ and showing that the integral corresponding to each term is zero, i.e. $\int_{-\infty}^\infty\,\dd\omega\,\ee^{\ii\,\omega\,(t-t')}\,\wt g_\fb(\omega)^n=0$ for $n\geq 1$. Because of the finite feedback delay time, the feedback filter function $\wt g_\fb(\omega)$ contains a phase term $\ee^{\ii\,\omega\,\tau_\fb}$, such that $\wt g_\fb(\omega)=\wt g_{\fb,0}(\omega)\ \ee^{\ii\,\omega\,\tau_\fb}$, where $\wt g_{\fb,0}(\omega)$ describes the feedback in the limit of zero delay. Hence the previous integral becomes $\int_{-\infty}^{\infty}\dd\omega\ 
\ee^{\ii\,\omega\pt{t-t'+n\,\tau_\fb}}\
\wt g_{\fb,0}(\omega)^n$.
When $\abs{t-t'}\leq\tau_\fb$, then  $t-t'+n\,\tau_\fb\geq 0$, $\forall n\geq 1$, so that the 
exponential term, in the complex plane, decays for increasing values of the imaginary part of the complex argument. This implies that the previous integral can be evaluated as the integral along the curve in the complex plane made by the x-axis and the half circle on the upper half plane, in the limit of infinite radius of the half circle. 
Since, in this region, $\wt g_{\fb,0}(\omega)$ [and hence also $\wt g_{\fb,0}(\omega)^n$] is analytic, then the integral is zero.

\subsubsection{Power spectrum of the in-loop field}\label{Sec.inloop}

\begin{figure}[t!]
\includegraphics[width=8.5cm]{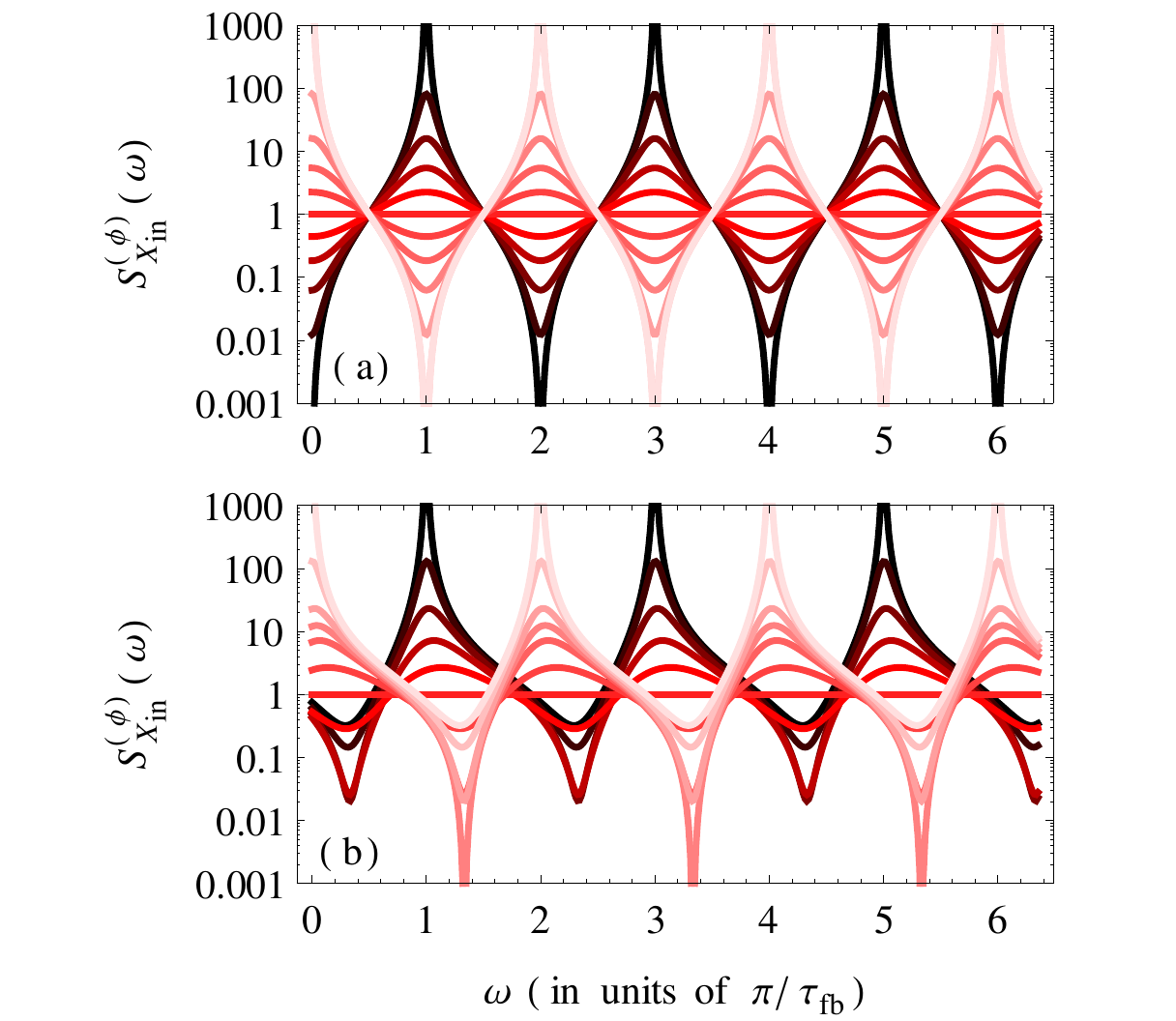}
\caption{Power spectrum of the in--loop field $S_{X_\inn\al{\phi}(\omega)}$ evaluated for the filter function defined in Eq.~\rp{gfbflat} (with $\phi_\fb=0$) and for (a) $\phi=\theta_\fb$ and (b) $\phi=\theta_\fb-\pi/3$.
Lines from 
dark to light red
correspond to values of $\bar g_\fb$ which range from $\bar g_\fb=-1/(2\sqrt{\eta}\,\cos\theta_\fb)$ to $\bar g_\fb=1/(2\sqrt{\eta}\,\cos\theta_\fb)$, and the horizontal red solid line corresponds to $\bar g_\fb=0$.
Both spectra are found under the same condition of Fig.~\ref{lambda0}.
In (a) we observe zeros of the power spectrum in correspondence with the minima of the photocurrent. In (b) the minima are shifted and found for a value of $\bar g_\fb$ different from that corresponding to the minima of the photocurrent.
}\label{SXin0}
\end{figure}

Let us now discuss the properties of the fluctuations of the in-loop field. In particular we show that the fluctuations of a specific in-loop field quadrature $X_{\rm in}\al{\phi}(\omega)$ can be fully suppressed at specific frequencies by destructive interference, when the detection efficiency is perfect ($\eta=1$). Differently from standard squashing~\cite{Shapiro87} (discussed in Sec.~\ref{squashing}) where the fluctuations of the detected quadrature can be suppressed in the limit of infinite negative gain, here
we show that the fluctuations of a quadrature different from the detected one can be suppressed
at finite feedback gain.

Specifically, we study here the power spectrum of a generic quadrature $X_{\rm in}\al{\phi}(\omega)$ with phase $\phi$ (that can be also different form the phase of the detected quadrature $\theta_\fb$).
Using the expressions for the photocurrent and for the in-loop operators in Eqs.~\rp{ifree} and \rp{ain}, respectively, we find
\begin{eqnarray}
\wt X_{\rm in}\al{\phi}(\omega)&=&
\wt X_{\rm in}^{\circ\, \pt{\phi}}(\omega)+2\,\cos\pt{\phi}\ \wt g_\fb(\omega)\ \lambda(\omega)\ \wt i^\circ(\omega)\ ,
\end{eqnarray}
so that the corresponding power spectrum, defined by the relation $\delta(\omega+\omega')\ S_{X_{\rm in}\al{\phi}}(\omega)=\av{\wt X_{\rm in}\al{\phi}(\omega)\ \wt X_{\rm in}\al{\phi}(\omega')}$
takes the form
\begin{eqnarray}
S_{X_{\rm in}\al{\phi}}(\omega)&=&
\abs{1+2\sqrt{\eta}\,\cos(\phi)\,\wt g_\fb(\omega)\ \lambda(\omega)\ \ee^{\ii\pt{\phi-\theta_\fb }}}^2
\nn\\&&
+4\,(1-\eta)\,\cos\pt{\phi}^2\,\abs{\wt g_\fb(\omega)}^2\,\abs{\lambda(\omega)}^2\ .
\end{eqnarray}
We note that for perfect photodetection $\eta=1$,
this expression reduces to
\begin{eqnarray}
S_{X_{\rm in}\al{\phi}}(\omega)\Bigl|_{\eta=1}&=&
\abs{1+2\,\cos(\phi)\,\wt g_\fb(\omega)\ \lambda(\omega)\ \ee^{\ii\pt{\phi-\theta_\fb }}}^2,
\end{eqnarray}
which is a coherent superposition of two terms. The first term corresponds to the fluctuations of a free field and the second one is due to the feedback. In particular the feedback term can be 
adjusted in order to realize perfectly destructive interference at a specific frequency. This effect is described by Fig.~\ref{SXin0} which shows the suppression of the in--loop fluctuations also for a field quadrature different from the detected one [see plot (b)].

We finally note that this effect cannot be observed directly.
In fact, as shown in Ref.~\cite{Shapiro87}, the reduced in-loop fluctuations cannot be extracted out of the loop using, for example, a beam splitter. Rather, the out-of-loop field
extracted with a beam splitter always exhibits classical super-shot-noise fluctuations~\cite{Shapiro87} (see also Sec.~\ref{ouofloop} below).
However,
the modified in-loop fluctuations can be indirectly probed by measuring their effects on an additional system which interacts
with the in-loop field~\cite{Yamamoto86,Shapiro87,Wiseman98}. Specifically, it has been recently shown~\cite{Rossi} (see also Sec.~\ref{Sec.cooling}) that the interference discussed above can be used to suppress certain scattering processes
in an optomechanical system hence enhancing the cooling efficiency.

\section{Feedback--controlled light with an optical cavity}\label{SecCav}

\begin{figure}[t!]
\includegraphics[width=8.5cm]{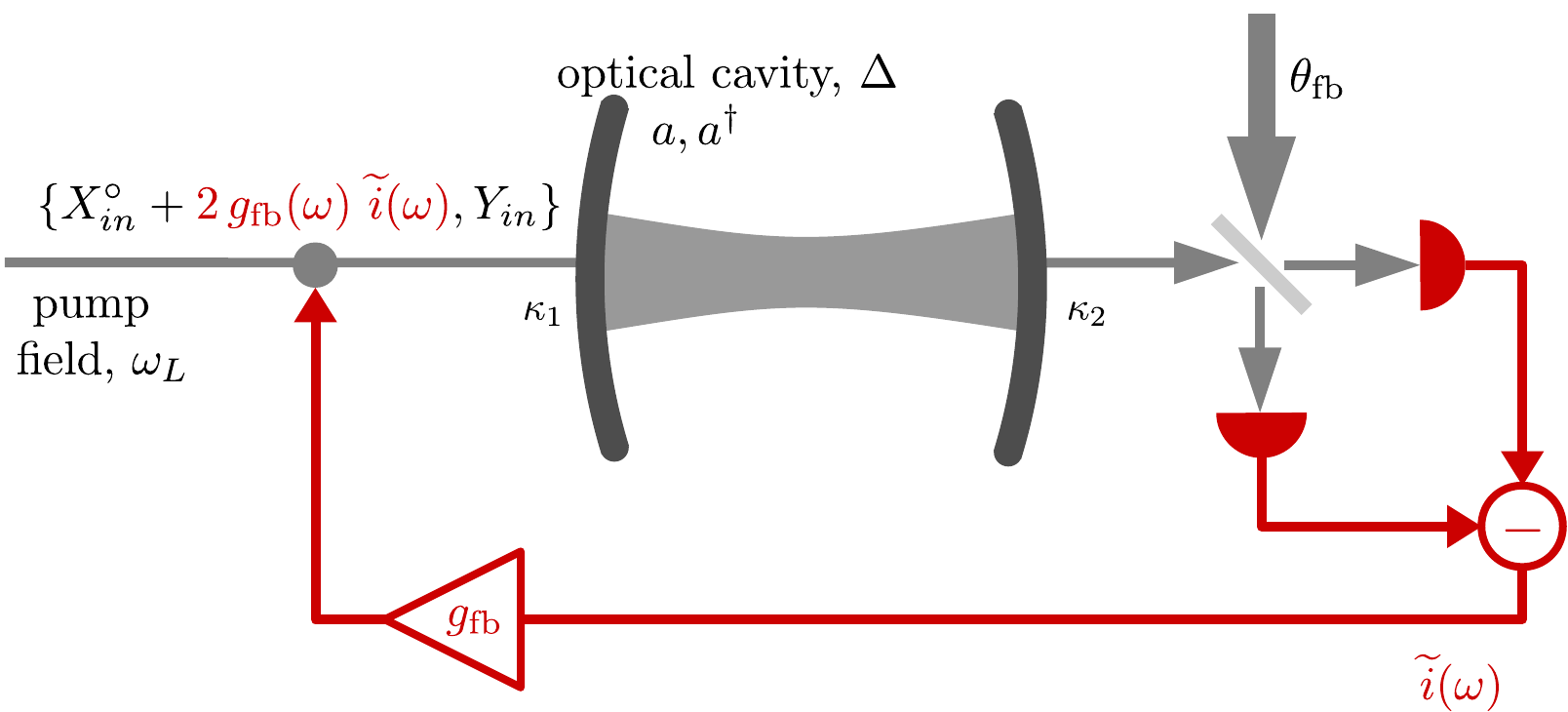}
\caption{The feedback loop: a quadrature at phase $\theta_\fb$ of the field transmitted through a Fabry-P\'erot optical cavity (detuned by $\Delta$ from the input field, and with dissipation rates $\kappa_1$ and $\kappa_2$) is detected, and the corresponding photocurrent is used to modulate the input amplitude $X_\inn$. In this case the feedback can be closed also by measuring the reflected field.
}\label{schema1}
\end{figure}

Here we study the effect of feedback on the field of a mode of an empty optical cavity placed inside the feedback loop, as shown in Fig.~\ref{schema1}~\cite{Sheard}. 
We will show that the steady state cavity field is in a classical thermal squeezed state.
Moreover, we discuss how a cavity within the feedback loop exhibits a modified susceptibility with a modified cavity decay rate, which can be controlled via the feedback parameters.
These and other results will be useful for the understanding of the in-loop optomechanical dynamics discussed in the next Section.

\subsection{The model}\label{modelcav}

Since the effect of the feedback loop depends upon the phase of the detected field, it is useful to explicitly include the phase difference between driving, cavity and output fields in the equations for the system operators.
In particular, we consider a resonant mode of an optical cavity at frequency $\omega_c$ and with amplitude decay rate $\kappa$, which is driven by a field at frequency $\omega_L=\omega_c-\Delta$, so that the cavity susceptibility is given by
\begin{eqnarray}\label{chic}
\chi_c(\omega)=\frac{1}{\kappa+\ii\pt{\Delta-\omega}} \ .
\end{eqnarray}
We analyse the case of a Fabry--P\'erot configuration with two mirrors with corresponding decay rates $\kappa_1$ and $\kappa_2$, and include also additional dissipation due to, for example, internal losses or absorption at rate $\kappa'$ (such that $\kappa=\kappa_1+\kappa_2+\kappa'$).
The annihilation operator for the cavity field fluctuations in Fourier space, $\wt a(\omega)$, fulfils the standard quantum Langevin equation
\begin{eqnarray}\label{aomega}
-\pq{\kappa+\ii\pt{\Delta-\omega}}\,\wt a(\omega)+
\ee^{-\ii\,\phi_c}\ \sqrt{2\,\kappa}\ \wt a_{\inn,\rm tot}(\omega)=0\ ,
\end{eqnarray}
where $\phi_c$ is the phase difference between the input and cavity fields, defined by the relation
\begin{eqnarray}\label{phic}
\ee^{\ii\phi_c}&=&\frac{\kappa-\ii\,\Delta}{\sqrt{\kappa^2+\Delta^2}}\ ,
\end{eqnarray}
and where
we have included the total input noise operator which can be decomposed in terms of the operators corresponding to the individual decay channels as
\begin{eqnarray}\label{aintot}
\wt a_{\inn,\rm tot}(\omega)=\frac{\sqrt{\kappa_1}\,\wt a_{\rm in,1}(\omega)+\sqrt{\kappa_2}\,\wt a_{\rm in,2}(\omega)+\sqrt{\kappa'}\,\wt a'_{\rm in}(\omega)}{\sqrt{\kappa}}\ .
\end{eqnarray}
The input noise operator without feedback $\wt a_{\inn,\rm tot}^\circ(\omega)$ describes vacuum fluctuations according to $\av{\wt a_{\inn,\rm tot}^\circ(\omega)\ \wt a_{\inn,\rm tot}^{\circ\,\dagger}(\omega')}=\delta(\omega+\omega')$ and $\av{\wt a_{\inn,\rm tot}^\circ(\omega)\ \wt a_{\inn,\rm tot}^\circ(\omega')}=0$ (similar expressions are valid also for the noise operators of each noise channel).
In particular, here we assume that the driving  field acts on the first mirror, so that
the input operator of the first mirror is modulated by the feedback according to the relation
\begin{eqnarray}\label{ain1}
\wt a_{\rm in,1}(\omega)=\wt a_{\rm in,1}^\circ(\omega)+\wt g_\fb(\omega)\ \wt i(\omega)\ .
\end{eqnarray}
The corresponding input--output relations that relate the cavity output to the cavity and input noise operators are
\begin{eqnarray}\label{aomega0}
\wt a_{\rm out,j}(\omega)\ \ee^{\ii\,\phi_{\rm out,j}}=\sqrt{2\,\kappa_j}\ \wt a(\omega)\ \ee^{\ii\,\phi_c}-\wt a_{\rm in,j}(\omega)\ ,
\end{eqnarray}
for $j=1,2$, and where we have introduced the phase difference, $\phi_{\out,j}$, between the input of the first mirror and the $j$-th output field. They are explicitly given by $\phi_{\out,2}=\phi_c$ and $\phi_{\out,1}=\phi_c+\phi_c'$, where the additional phase of the reflected field is defined by the relation $\ee^{\ii\,\phi_c'}=\pt{2\,\kappa_1-\kappa-\ii\,\Delta}/\sqrt{(2\,\kappa_1-\kappa)^2+\Delta^2}$.
Using these expressions it is now possible to analyse the dynamics of an in-loop cavity.

\subsection{Feedback photocurrent with a cavity}\label{S_i_cav}

Let us first study the feedback photocurrent.
As in the previous section, here we assume that we detect a quadrature at phase $\theta_\fb$, of one of the two outputs, $\wt X_{\out,\fb}\al{\theta_\fb}(\omega)=\ee^{-\ii\,\theta_\fb}\ \wt a_{\out,\fb}(\omega)+\ee^{\ii\,\theta_\fb}\ \wt a_{\out,\fb}\da(\omega)$, where
$\wt a_{\out,\fb}(\omega)=\wt a_{\rm out,1}(\omega)$ if the feedback is closed by measuring the reflected field, while $\wt a_{\out,\fb}(\omega)=\wt a_{\rm out,2}(\omega)$ when the feedback is closed in transmission (as in the specific case depicted in Fig.~\ref{schema1}).
Then, the photocurrent takes the form
\begin{eqnarray}\label{wtitheta}
\wt i(\omega)&=&\sqrt{\eta}\ \wt X_{\out,\fb}\al{\theta_\fb}(\omega)+\sqrt{1-\eta}\ \wt X_v(\omega)\ ,
\end{eqnarray}
which can be equivalently expressed in terms of the photocurrent without feedback $\wt i^\circ(\omega)$
(the power spectrum of which, also in this case, is equal to one)  as
\begin{eqnarray}\label{icav}
\wt i(\omega)&=&
\wt i^\circ(\omega)+2\,\sqrt{\eta}\,\zeta_\fb\al{\bar\theta_\fb}(\omega)\ \wt g_\fb(\omega)\ \wt i(\omega)\ .
\end{eqnarray}
In this expression we have introduced the cavity response function $\zeta_\fb\al{\bar\theta_\fb}(\omega)$ that describes how input amplitude fluctuations are transferred to the output (i.e. $\wt X_{\out,\fb}\al{\theta_\fb}(\omega)=\zeta_\fb\al{\bar\theta_\fb}(\omega)\ \wt X_{\rm in,1}(\omega)+\cdots$ where the dots stand for contributions due to other input noise operators). It is given by
\begin{eqnarray}\label{zetafb_2}
\zeta_\fb\al{\bar\theta_\fb}(\omega)=
2\,\sqrt{\kappa_\fb\,\kappa_1}\ \zeta_c\al{\bar\theta_\fb}(\omega)-
\lpg{\mat{cl}{
0
 &\text{\ \ in transmission}\\
\cos(\bar\theta_\fb)
&\text{\ \ in reflection}}
}
\end{eqnarray}
where we have introduced the cavity transfer function
\begin{eqnarray}\label{zetac}
\zeta_c\al{\bar\theta_\fb}(\omega)&=&
\frac{
\ee^{-\ii\,\bar\theta_\fb}\ \chi_c(\omega)+\ee^{\ii\,\bar\theta_\fb}\ \chi_c(-\omega)^*
}{2}\ ,
\end{eqnarray}
the feedback phase $\bar\theta_\fb$, which includes also the phase difference between input and output
\begin{eqnarray}\label{theta}
\bar\theta_\fb=
\lpg{
\mat{cl}{
\theta_\fb+\phi_{\rm out,2}  &\text{\ \ in transmission}\\
\theta_\fb+\phi_{\rm out,1}  &\text{\ \ in reflection}}
}\ ,
\end{eqnarray}
and the decay rate, $\kappa_\fb$, of the mirror corresponding to the detected output
\begin{eqnarray}\label{kfb}
\kappa_\fb=\lpg{\mat{cl}{
\kappa_2 &\text{\ \ in transmission}\\
\kappa_1&\text{\ \ in reflection}}
}\ .
\end{eqnarray}
%
%
We note that the term $\cos(\bar\theta_\fb)$ in Eq.~\rp{zetafb_2}, which is relevant for the feedback closed in reflection, is due to the component of the input field that is directly reflected by the first mirror, while the term proportional to cavity transfer function $\zeta_c\al{\bar\theta_\fb}(\omega)$ accounts for the component of the input field that is filtered by the cavity.

\begin{figure}[t!]
\includegraphics[width=8.cm]{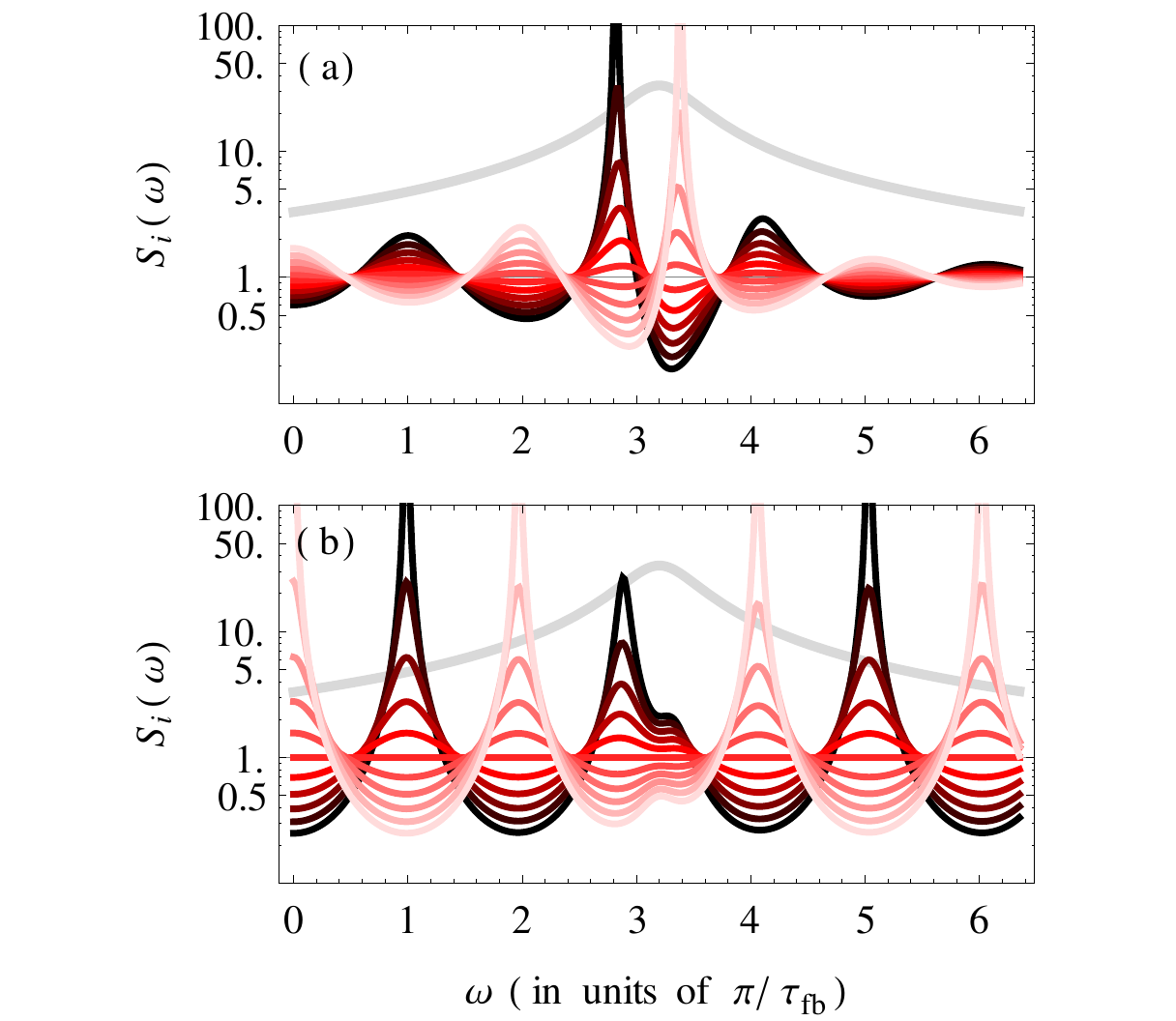}
\caption{Photocurrent power spectrum normalized such that the power spectrum for a coherent field is equal to one, and evaluated for the flat filter function $\wt g_\fb(\omega)$ defined in Eq.~\rp{gfbflat} (with $\phi_\fb=0$), with perfect detection efficiency ($\eta=1$), and including a symmetric cavity ($\kappa_1=\kappa_2$). Different colours correspond to different values of $\bar g_\fb$ with dark to light red
corresponding to increasing values in the range of feedback stability as defined in Sec.~\ref{stabiliycav},
and the horizontal solid red 
line corresponds to $\bar g_\fb=0$.
The thick grey line indicates the position of the cavity with $\kappa=1/\tau_\fb$ and $\Delta=10/\tau_\fb$. Plot (a) corresponds to the feedback closed in transmission and (b) in reflection.
}\label{lambda1}
\end{figure}

In order to describe compactly both configurations (i.e. feedback closed in transmission and in reflection) and to simplify various expressions in the next sections, it is useful to introduce the following notation.
We define a modified feedback function
\begin{eqnarray}\label{hfb}
\wt h_\fb(\omega)=\lpg{\mat{cl}{
\wt g_\fb(\omega) &\text{\ \ in transmission}\\
\frac{\wt g_\fb(\omega)}{1+2\sqrt{\eta}\,\wt g_\fb(\omega)\,\cos(\bar\theta_\fb)} &\text{\ \ in reflection}}
}\ ,
\end{eqnarray}
which, when the feedback is closed in reflection, accounts for the effect of the component of the input field directly reflected from the first mirror.
Thereby, using this definition in Eq.~\rp{icav} we find
\begin{eqnarray}\label{gi}
\wt g_\fb(\omega)\ \wt i(\omega)=\wt h_\fb(\omega)\ \lambda_{c,\fb}(\omega)\ \wt i^\circ(\omega)
\end{eqnarray}
where we have introduced the squashing factor
\begin{eqnarray}\label{lambda_fb}
\lambda_{c,\fb}(\omega)=\frac{1}{1-2\,\mu_\fb(\omega)\,\zeta_c\al{\bar\theta_\fb}(\omega)}\ ,
\end{eqnarray}
which includes the feedback transfer function
\begin{eqnarray}\label{mufb}
\mu_\fb(\omega)=2\sqrt{\kappa_\fb\,\kappa_1\,\eta}\  \wt h_\fb(\omega)\ .
\end{eqnarray}
The corresponding
photocurrent power spectrum is given by
\begin{eqnarray}\label{Sicav}
S_i(\omega)=\abs{\frac{\wt h_\fb(\omega)}{\wt g_\fb(\omega)}}^2\ \abs{\lambda_{c,\fb}(\omega)}^2\ ,
\end{eqnarray}
and it is reported in Fig~\ref{lambda1}.
In this case the feedback signal is filtered not only by the electronic filter function $\wt g_\fb(\omega)$ as in the previous section, but also by the cavity, through the cavity transfer function in Eq.~\rp{zetafb_2} [or equivalently Eq.~\rp{zetac}].
As shown in Fig~\ref{lambda1} (a), when the loop is closed in transmission, the feedback is effective only within the cavity linewidth, while it is strongly suppressed away form the cavity resonance. In reflection, instead, the feedback is relevant for all frequencies [see Fig~\ref{lambda1} (b)], due to the component of the field that is directly reflected by the first mirror. The presence of the cavity affects
the feedback response around
the range of frequencies covered by the cavity.

\subsubsection{Feedback stability}\label{stabiliycav}

\begin{figure}[t!]
\includegraphics[width=8.5cm]{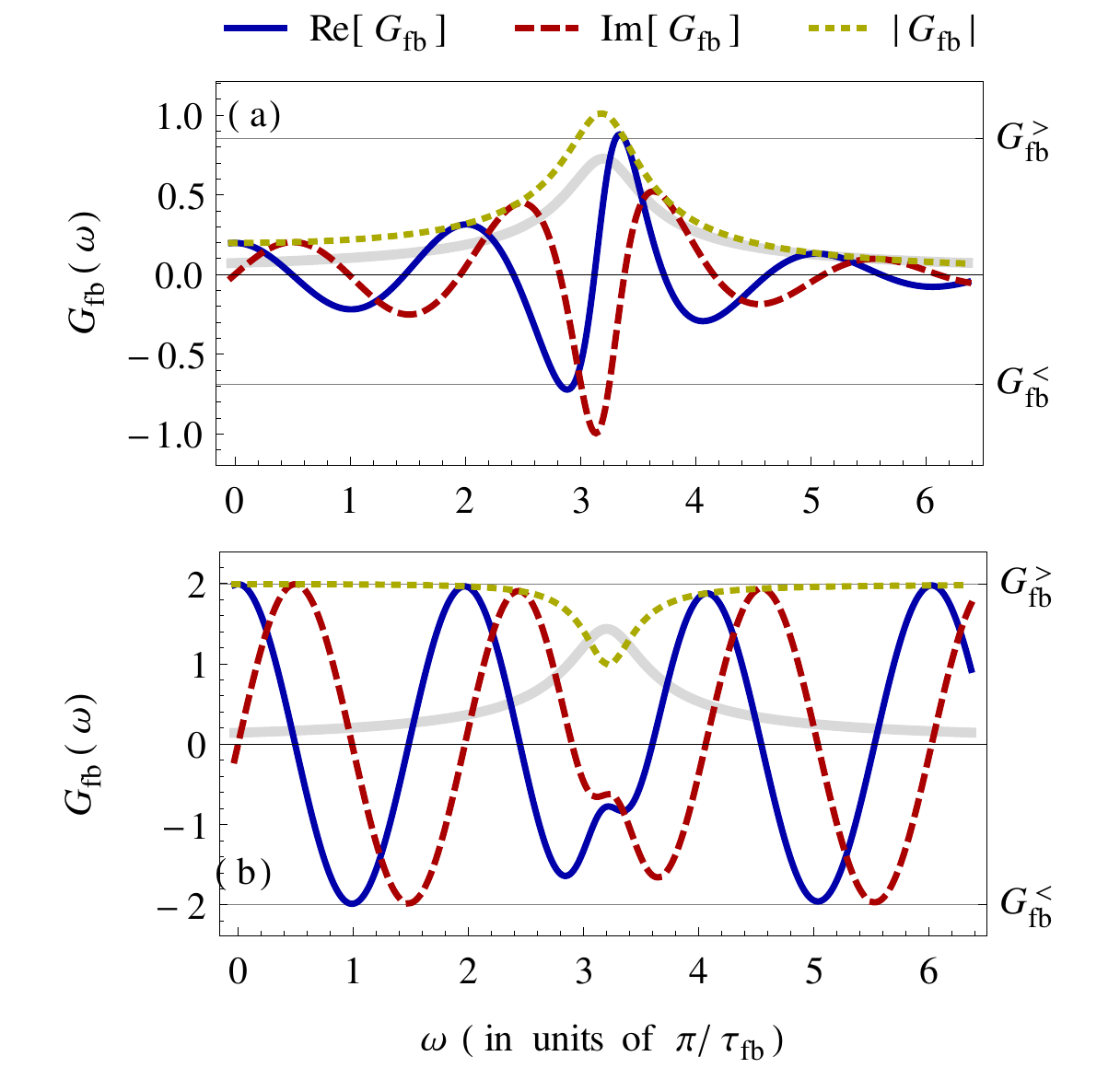}
\caption{Total feedback transfer function $G_\fb(\omega)$ (see Sec.~\ref{stabiliycav}) evaluated for the same parameters and feedback configurations of Fig.~\ref{lambda1} [(a) is in transmission and (b) in reflection], but with $\bar g_\fb=1$.
The horizontal lines indicate the values of $G_\fb^>$ and $G_\fb^<$ introduced in Sec.~\ref{stabiliycav}, such that the feedback is stable when $1/G_\fb^><\bar g_\fb<1/G_\fb^<$. The blue, solid lines are the real part ${\rm Re}[G_\fb(\omega)]$, the red, dashed lines the imaginary part ${\rm Im}[G_\fb(\omega)]$, and the yellow, dotted lines are the absolute value $|G_\fb(\omega)|$. The thick grey line indicates the position of the cavity with $\kappa=1/\tau_\fb$ and $\Delta=10/\tau_\fb$.
}\label{Gfbcav}
\end{figure}

The photocurrent power spectrum in Eq.~\rp{Sicav} can be expressed in terms of the total feedback response function $G_\fb(\omega)=\frac{\wt g_\fb(\omega)}{\wt h_\fb(\omega)}\pq{1-2\,\mu_\fb(\omega)\,\zeta_c\al{\bar\theta_\fb}(\omega)}-1$ as
$S_i(\omega)=\abs{1-G_\fb(\omega)}^{-2}$. The feedback becomes unstable when $G_\fb(\omega)=1$. In particular, the real and imaginary parts of $G_\fb(\omega)$ oscillate between negative and positive values, as in Fig.~\ref{Gfbcav}, so that the feedback is stable if the real part, evaluated for the discrete set of frequencies $\pg{\omega_i}$ where the imaginary part is zero (such that ${\rm Im}\pq{G_\fb(\omega_i)}=0$) is smaller then one, i.e. ${\rm Re}\pq{G_\fb(\omega_i)}<1$. Thus, in the case of the flat feedback function~\rp{gfbflat}, in order to determine the stability conditions in terms of the values of $\bar g_\fb$,
we can introduce the maximum and minimum of ${\rm Re}\pq{G_\fb(\omega_i)}$ evaluated for $\bar g_\fb=1$, that is $G_\fb\al{>}={\rm max}\pg{{\rm Re}\pq{G_\fb(\omega_i)\bigr|_{\bar g_\fb=1}}}$ and $G_\fb\al{<}={\rm min}\pg{{\rm Re}\pq{G_\fb(\omega_i)\bigr|_{\bar g\fb=1}}}$, and state that the feedback is stable in the range ${1}/{G_\fb\al{>}}<\bar g_\fb<{1}/{G_\fb\al{<}}$ (see Fig.~\ref{Gfbcav}).

\subsection{The unused (out--of--loop) output field}\label{ouofloop}

\begin{figure}[t!]
\includegraphics[width=8.5cm]{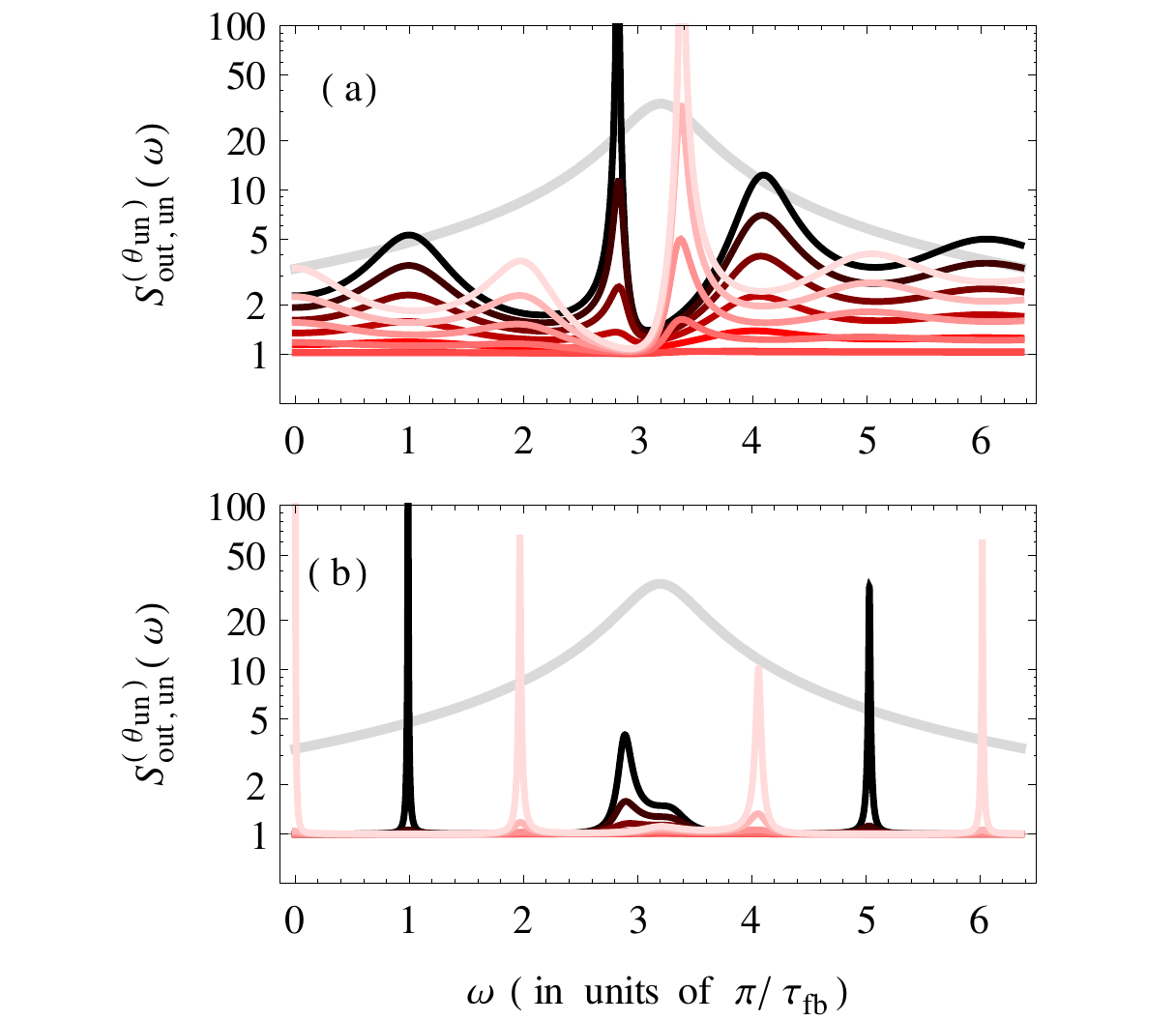}
\caption{Power spectrum of the quadrature at phase $\theta_\un=\theta_\fb$ of the field lost by the cavity from the unused output port. It is evaluated for the same parameters and feedback configurations of Fig.~\ref{lambda1}. In (a) the feebdack is closed in transmission so this plot corresponds to the reflected field. In (b) instead the feedback is closed in reflection and this plot shows the fluctuations of the transmitted field.
The unused (out--of--loop) output field 
always exhibits fluctuations enhanced with respect to the vacuum noise level (which is here set to one).
}\label{Soutfcav}
\end{figure}

In this section we study the properties of the light at the cavity output that is not used for the feedback, and we show that this light always exhibits super--shot--noise fluctuations.
In particular,
a quadrature at phase $\theta_\un$ of the out--of--loop field at the output of the unused cavity mirror, can be
expressed as
\begin{eqnarray}\label{Xoutfcav}
\wt X_{\out,\un}\al{\theta_\un}(\omega)
&=&\wt X_{\out,\un}^{\circ\, \pt{\theta_\un}}(\omega)+2\,\zeta_\un\al{\theta_\un}(\omega)\ \wt h_\fb(\omega)\ \lambda_{c,\fb}(\omega)
\ \wt i^\circ(\omega)\ ,
\nn\\
\end{eqnarray}
where, as usual, the symbol $^\circ$ indicates operators and quantities with no feedback, and
where $\zeta_\un\al{\theta_\un}(\omega)$ is the function that describes how input noise fluctuations from the first mirror are transferred to the non-detected output, such that $\wt X_{\out,\un}\al{\theta_\un}(\omega)=\zeta_\un\al{\theta_\un}(\omega)\ \wt X_{\rm in,1}(\omega)+\cdots$, with the dots indicating terms proportional to other input noise operators. Its explicit form is analogous to the one of the transfer function in Eq.~\rp{zetafb_2}, but with the roles of the parameters of the transmitted and reflected fields exchanged.

In the case of an empty cavity, the power spectra of the two output field quadratures $S_{\out,x}^{\circ\,(\theta_x)}(\omega)\ \delta(\omega+\omega')=\av{\wt X_{\out,x}^{\circ\, \pt{\theta_x}}(\omega)\ \wt X_{\out,x}^{\circ\, \pt{\theta_x}}(\omega')}$ [where $x\in\pg{\fb,\un}$ distinguishes the output that is used for the feedback ($x=\fb$) from the unused output ($x=\un$)],
when no feedback is applied, are equal to the vacuum noise that is here set to one, i.e.
$S_{\out,x}^{\circ\,(\theta_x)}(\omega)=1$. Moreover the cross power spectrum $S_{\out,\fb-\un}^{\circ\,(\theta_\fb,\theta_\un)}(\omega)\ \delta(\omega+\omega')=\av{\wt X_{\out,\fb}^{\circ\, \pt{\theta_\fb}}(\omega)\ \wt X_{\out,\un}^{\circ\, \pt{\theta_\un}}(\omega')}$ is zero. Thereby, we find that the power spectrum of the unused output in the presence of feedback is equal to
\begin{eqnarray}
S_{\out,\un}^{(\theta_\un)}(\omega)=1+\abs{2\,\zeta_\un\al{\theta_\un}(\omega)\ \wt h_\fb(\omega)\ \lambda_{\om,\fb}(\omega)}^2\ ,
\end{eqnarray}
which is always larger then the vacuum noise level (see Fig.~\ref{Soutfcav}). This shows that in--loop reduced fluctuations cannot be extracted out of the loop and hence do not correspond to actual squeezing~\cite{Shapiro87}.

\subsection{The cavity field}\label{cavcomm}

Here we study the properties of the cavity field described by Eq.~\rp{aomega}.
%
%
We first note that while in general in--loop fields do not fulfil standard bosonic commutation relations, the operators for the cavity mode do. This can be explicitly seen by constructing the commutators as the inverse Fourier transform of the corresponding expression in frequency which are found from Eq.~\rp{aomega}. Then, the integral of the inverse Fourier transform can be computed exploiting the analytic properties of $\wt g_\fb(\omega)$, and $\chi_c(\omega)$, and one finds  $\pq{a(t),a\da(t)}=1$ and $\pq{a(t),a(t)}=0$.


The cavity dynamics is Gaussian, so that  the cavity steady state is fully characterized by the correlation matrix of the field operators 
(note that a feedback loop closed using a high--pass response function does not affect the average field amplitude).
The correlation matrix can be expressed in terms of the vector of operators $\va(t)=\pt{a(t),a\da(t)}^T$ as
$\CC=\av{\va(t)\ \va(t)^T}$ with elements $\pg{\CC}_{j,k}=\av{\pg{\va(t)}_j\ \pg{\va(t)}_k}$. In particular the steady state $\CC_{st}$ can be found as the integral of the spectral density matrix $\SSS_\va(\omega)$ which is defined in terms of the
vector of cavity operators in Fourier space $\wt\va(\omega)=\pt{\wt a(\omega),\wt a\da(\omega)}^T$ according to the relation $\av{\wt\va(\omega)\ \wt\va(\omega')^T}=\SSS_\va(\omega)\ \delta(\omega+\omega')$, so that
\begin{eqnarray}\label{CCst0}
\CC_{st}=\frac{1}{2\,\pi}\int_{-\infty}^\infty\ \dd\omega\ \SSS_\va(\omega)\ .
\end{eqnarray}
Using the definition for the field operators and the photocurrent introduced in Secs.~\ref{modelcav} and \ref{S_i_cav} we find
\begin{eqnarray}\label{SSSva}
\SSS_\va(\omega)&=&2\,\kappa\,\abs{\chi_c(\omega)}^2\, \mm{0& 1}{0& 0}
\\&&\hspace{-1cm}
+2\,\kappa_1\,\abs{\wt h_\fb(\omega)\ \lambda_{c,\fb}(\omega)}^2
\nn\\&&\hspace{-0.5cm}\times
\mm{
\chi_c(\omega)\,\chi_c(-\omega)\,\ee^{-2\,\ii\,\phi_c} & \abs{\chi_c(\omega)}^2
}{
\abs{\chi_c(-\omega)}^2 & \chi_c(\omega)^*\,\chi_c(-\omega)^*\,\ee^{2\,\ii\,\phi_c}
}
\nn\\&&\hspace{-1cm}
+\ee^{-\ii\,\bar\theta_\fb}\ \chi_c(\omega)\ \mu_\fb(\omega)\ \lambda_{c,\fb}(\omega)
\mm{
0 & \chi_c(\omega)
}{
0 & \chi_c(-\omega)^*\,\ee^{2\,\ii\,\phi_c}
}
\nn\\&&\hspace{-1cm}
+\ee^{\ii\,\bar\theta_\fb}\ \chi_c(\omega)^*\ \mu_\fb(\omega)^*\ \lambda_{c,\fb}(\omega)^*
\mm{
\chi_c(-\omega)\,\ee^{-2\,\ii\,\phi_c} & \chi_c(\omega)^*
}{
0 & 0
}\ .
\nn
\end{eqnarray}
Using the fact that the functions $\chi_c(\omega)^2$, $\chi_c(\omega)\, \chi_c(-\omega)^*$ and $\mu_\fb(\omega)$ are analytic in the upper half complex plane,
one can show that when performing the integral in Eq.~\rp{CCst0}, the last two terms in Eq.~\rp{SSSva} give no contribution, i.e. $\int\,\dd\omega\ \chi_c(\omega)^2\ \mu_\fb(\omega)\ \lambda_c(\omega)
=
\int\,\dd\omega\ \chi_c(\omega)\, \chi_c(-\omega)^*\ \mu_\fb(\omega)\ \lambda_c(\omega)=0$.
Thereby we find that the stationary correlation matrix takes the form
\begin{eqnarray}\label{Cst}
\CC_{st}=\mm{m_{st}& n_{st}+1}{n_{st}& m_{st}^*},
\end{eqnarray}
where
\begin{eqnarray}\label{nstmst}
n_{st}&=&\frac{2\,\kappa_1}{2\,\pi}\int_{-\infty}^\infty\dd\omega\ \abs{\wt h_\fb(\omega)\ \lambda_{c,\fb}(\omega)\ \chi_c(\omega)}^2,
\nn\\
m_{st}&=&\frac{2\,\kappa_1}{2\,\pi}\ \ee^{-2\,\ii\,\phi_c}
\nn\\&&\times
\int_{-\infty}^\infty\dd\omega\ \abs{\wt h_\fb(\omega)\ \lambda_{c,\fb}(\omega)}^2 \ \chi_c(\omega) \ \chi_c(-\omega),
\end{eqnarray}
and we have used the result $\frac{2\,\kappa}{2\,\pi}\int_{-\infty}^\infty\dd\omega\ \abs{\chi_c(\omega)}^2=1$.
We finally highlight that Eq.~\rp{nstmst} implies $n_{st}>\abs{m_{st}}$, which, in turn, implies that no quadrature has a variance below the vacuum noise level, so that $\CC_{st}$ describes a classical squeezed thermal state~\cite{GardinerZoller}.

\subsection{The effective cavity susceptibility}\label{Sec:EffectiveCavity}

Here we study when it is meaningful to define an effective susceptibility which accounts for the modifications of the cavity dynamics due to the feedback.

\subsubsection{Cavity response to an additional input seed}

In Refs.~\cite{Rossi,Kralj,Rossi2} we have shown that an operational way to determine how the cavity susceptibility is modified by the feedback is to look at the cavity response to an additional driving probe seed. In order to achieve this while the feedback is active without affecting the feedback itself, we have added an additional tone, at frequency $\omega_L+\nu$, to the pump field (that is at frequency $\omega_L$), and  with amplitude $\alpha_s$ much smaller than the pump, but at the same time much larger than the fluctuations. One can then look at the response (the photocurrent) at the frequency of the probe $\nu$ which in turn is scanned around the pump frequency. Specifically we have considered the input noise operator of the form
\begin{eqnarray}
\wt a_{\rm in,1}(\omega)=\wt a_{\rm in,1}^\circ(\omega)+\wt g_\fb(\omega)\,\wt i(\omega)+\alpha_s\,\delta(\omega-\nu)\ .
\end{eqnarray}
When considering the feedback in transmission as in Refs.~\cite{Rossi,Kralj,Rossi2} the photocurrent is therefore given by
\begin{eqnarray}
\wt i(\omega)&=&\lambda_{c,\fb}(\omega)\,\wt i^\circ(\omega)
+2\,\sqrt{\kappa_2\,\kappa_1\,\eta}\,\alpha_s\,\lambda_c(\omega)
\nn\\&&\hspace{-1.5cm}\times
\pq{
\chi_c(\omega)\ \ee^{-\ii\,\theta_\fb}\,\delta(\omega-\nu)
+\chi_c(-\omega)^*\ \ee^{\ii\,\theta_\fb}\,\delta(\omega+\nu)
}\ ,
\end{eqnarray}
so that the corresponding power spectrum at frequency $\nu$ can be approximated as
\begin{eqnarray}
S_s(\nu)\simeq4\,\kappa_1\,\kappa_2\,\eta\,\alpha_s^2\abs{\lambda_{c,\fb}(\nu)}^2\ \abs{\chi_c(\nu)}^2\ ,
\end{eqnarray}
where we have neglected the vacuum light fluctuations under the assumption of sufficiently large $\alpha_s$.
This result indicates that
the system response is characterized by the effective cavity susceptibility
\begin{eqnarray}\label{chieff}
\chi_c^{\rm eff}(\omega)=\chi_c(\omega)\ \lambda_{c,\fb}(\omega)\ .
\end{eqnarray}

\subsubsection{Effective model}

Here we discuss when the effective susceptibility that we have identified above properly describes the cavity dynamics.

Including the equation for the feedback--modified input operator~\rp{ain1} into the equation for the cavity field~\rp{aomega},
makes explicit the dependence of the cavity field operator on the feedback photocurrent. In turn, the photocurrent depends on the cavity field and on the input noise operators themselves. In particular,
according to its definition in Eq.~\rp{wtitheta}, and the input--output relation~\rp{aomega0},  we find
\begin{eqnarray}
\wt g_\fb(\omega)\ \wt i(\omega)
&=&\wt h_\fb(\omega)
\lpg{
\sqrt{\eta}\pq{\sqrt{2\kappa_\fb}\wt X^{\pt{\bar\theta_\fb-\phi_c}}(\omega)-X_{\rm in,fb}^{\circ\,\pt{\bar\theta_\fb}}(\omega)}
}\nn\\&&\rpg{
+\sqrt{1-\eta}\wt X_v(\omega)
}\ .
\nn\\
\end{eqnarray}
with  $\bar\theta_\fb$ and $\wt h_\fb(\omega)$ defined in Eqs.~\rp{theta} and \rp{hfb} respectively.
Using this expression, the equations for the cavity field operators in Fourier space $\wt a(\omega)$ and $\wt a\da(\omega)\equiv\pg{\wt a(-\omega)}\da$ can be rewritten
as
\begin{eqnarray}\label{Model2}
&&-\pg{\wt\kappa_\eff(\omega)+\ii\pq{\wt\Delta_\eff(\omega)-\omega}}\,\wt a(\omega)
\\&&\hspace{0.2cm}
+\mu_\fb(\omega)\ \ee^{\ii\,\pt{\bar\theta_\fb-2\,\phi_c}}\,\wt a\da(\omega)
+ \sqrt{2\,\wt\kappa_\fb(\omega)}\ \wt a_{\rm in,\eff}(\omega)\ \ee^{-\ii\,\phi_c}=0\ ,
\nn\\
&&-\pg{\wt\kappa_\eff(-\omega)-\ii\pq{\wt\Delta_\eff(-\omega)+\omega}}\,\wt a\da(\omega)
\nn\\&&\hspace{0.2cm}
+\mu_\fb(\omega)\ \ee^{-\ii\,\pt{\bar\theta_\fb-2\,\phi_c}}\,\wt a(\omega)
+ \sqrt{2\,\wt\kappa_\fb(-\omega)}\ \wt a_{\rm in,\eff}\da(\omega)\ \ee^{\ii\,\phi_c}=0\ ,
\nn
\end{eqnarray}
%
%
with the feedback transfer function $\mu_\fb(\omega)$ introduced in Eq.~\rp{mufb} [note that it fulfils the relation $\mu_\fb(\omega)=\mu_\fb(-\omega)^*$], and
where we have introduced the effective frequency--dependent parameters
\begin{eqnarray}\label{Dkeffomega}
\wt\Delta_\eff(\omega)&=&\Delta-{\rm Im}\pq{\ee^{-\ii\bar\theta_\fb}\ \mu_\fb(\omega)}\ ,
\nn\\
\wt\kappa_\eff(\omega)&=&\kappa-{\rm Re}\pq{\ee^{-\ii\bar\theta_\fb}\ \mu_\fb(\omega)}\ ,
\end{eqnarray}
and the effective
input noise operator
 \begin{eqnarray}
\wt a_{\rm in,\eff}(\omega)&=&
\sqrt{\frac{\kappa}{\wt\kappa_\eff(\omega)}}\,\wt a^\circ_{\rm in,tot}(\omega)
\\&&\hspace{-1.5cm}
-\frac{\mu_\fb(\omega)}{2\,\sqrt{\wt\kappa_\eff(\omega)\,\kappa_\fb\,\eta}}\ \pq{
\sqrt{\eta}\,X_{\rm in,fb}^{\circ\,\pt{\bar\theta_\fb}}(\omega)
-\sqrt{1-\eta}\,\wt X_v(\omega)
}\ ,
\nn
\end{eqnarray}
with $\wt a^\circ_{\rm in,tot}(\omega)$ defined as in Eq.~\rp{aintot}.
This effective noise operator is characterized by the correlation functions $\av{\wt a_{\rm in,\eff}\da(\omega)\ \wt a_{\rm in,\eff}(\omega')}=\delta(\omega+\omega')\ \wt n_{\rm in}(-\omega)$, $\av{\wt a_{\rm in,\eff}(\omega)\ \wt a_{\rm in,\eff}\da(\omega')}=\delta(\omega+\omega')\pq{ \wt n_{\rm in}(\omega)+1}$, and $\av{\wt a_{\rm in,\eff}(\omega)\ \wt a_{\rm in,\eff}(\omega')}=\delta(\omega+\omega')\ \wt m_{\rm in}(\omega) $, with
\begin{eqnarray}
\wt n_{\rm in}(\omega)&=&
\frac{\abs{\mu_\fb(\omega)}^2}{4\,\wt\kappa_\eff(\omega)\,\kappa_\fb\,\eta}\ ,
\\
\wt m_{\rm in}(\omega)&=&\frac{1}{2\,\sqrt{\wt\kappa_\eff(\omega)\ \wt\kappa_\eff(-\omega)}}
\pq{
\frac{\abs{\mu_\fb(\omega)}^2}{2\,\kappa_\fb\,\eta}-\mu_\fb(\omega)^*\ \ee^{\ii\,\bar\theta_\fb}
}\ .
\nn
\end{eqnarray}
This operator is, however, an in-loop operator, and as such it does not fulfil standard bosonic commutation relations (see Sec.~\ref{comm}). In fact $\pq{\wt a_{\rm in,\eff}(\omega),\wt a_{\rm in,\eff}\da(\omega')}=\delta(\omega+\omega')$,
and $\pq{\wt a_{\rm in,\eff}(\omega),\wt a_{\rm in,\eff}(\omega')}=\delta(\omega+\omega')\ \ee^{\ii\,\pt{\bar\theta_\fb-2\,\phi_c}}\ {\rm Im}\pq{\mu_\fb(\omega)}/\sqrt{\wt\kappa_\eff(\omega)\ \wt\kappa_\eff(-\omega)}$.
We note, nevertheless, that as shown in Sec.~\ref{cavcomm} the cavity operators $a(t)$ and $a(t)\da$ are well defined bosonic operators.

The solution of Eq.~\rp{Model2} can be cast in the form
\begin{eqnarray}
\wt a(\omega)&=&\ee^{-\ii\,\phi_c}\ \chi_c^\eff(\omega)\ \wt f_{\rm in,c}(\omega),
\end{eqnarray}
with the total noise operator given by
\begin{eqnarray}\label{fcin}
\wt f_{\rm in,c}(\omega)&=&
\pq{
1- \chi_c(-\omega)^*\ \mu_\fb(\omega)\,\ee^{\ii\,\bar\theta_\fb}
}\,\sqrt{2\,\wt\kappa_\eff(\omega)}\ \wt a_{\rm in,\eff}(\omega)
\nn\\&&
+ \chi_c(-\omega)^*\ \mu_\fb(\omega)\,\ee^{\ii\,\bar\theta_\fb}\ \sqrt{2\,\wt\kappa_\eff(-\omega)}\ \wt a_{\rm in,\eff}\da(\omega),
\end{eqnarray}
which shows that the cavity field is proportional to the function $\chi_c^\eff(\omega)=\chi_c(\omega)\ \lambda_{c,\fb}(\omega)$, introduced in Eq.~\rp{chieff},
and this can justify
the interpretation of this function as the effective susceptibility of a feedback--controlled cavity.
Moreover it is interesting to note that the specific cavity field quadrature $\wt X\al{\bar\theta_\fb-\phi_c}(\omega)=\ee^{-\ii\,(\bar\theta_\fb-\phi_c)}\ \wt a(\omega)+\ee^{\ii\,(\bar\theta-\phi_c)}\ \wt a\da(\omega)$
at phase $\bar\theta_\fb-\phi_c$
[where $\bar\theta_\fb$ is related to the phase of the detected quadrature by Eq.~\rp{theta}],
takes the particularly simple form
\begin{eqnarray}\label{Xbarthetamphic}
\wt X\al{\bar\theta_\fb-\phi_c}(\omega)&=&
\chi_c^\eff(\omega)\
\sqrt{2\,\wt\kappa_\eff(\omega)}\ \wt a_{\rm in,\eff}(\omega)\ \ee^{-\ii\,\bar\theta_\fb}
\nn\\&&
+\ \chi_c^\eff(\omega)\
\sqrt{2\,\wt\kappa_\eff(-\omega)}\ \wt a_{\rm in,\eff}\da(\omega)\ \ee^{\ii\,\bar\theta_\fb},
\end{eqnarray}
which has the structure of a generic quadrature without feedback, that is $\wt X^{\circ\,\pt{\phi}}(\omega)=\sqrt{2\,\kappa}[\,\chi_c(\omega)\ \wt a_{\rm in,tot}^\circ(\omega)\ \ee^{-\ii(\phi+\phi_c)}+\chi_c(-\omega)^*\ \wt a_{\rm in,tot}^{\circ\,\dagger}(\omega)\ \ee^{\ii(\phi+\phi_c)}\,]$, but with the effective parameters in place of the original ones.
This implies that an additional system which is directly coupled to a quadrature operator at phase $\bar\theta_\fb-\phi_c$, via, for example, a Hamiltonian of the form $H_I\propto \hat s\ \wt X\al{\bar\theta_\fb-\phi_c}(\omega)$, where $\hat s$ is a generic operator of the additional system, would experience the effect of a modified cavity with susceptibility $\chi_\eff(\omega)$ and input noise operator $\wt a_{\rm in,\eff}(\omega)$.
%
This is, for example, the case in the experimental situation studied in Refs.~\cite{Rossi,Kralj,Rossi2} where the feedback is operated in transmission by measuring the output amplitude quadrature with $\theta_\fb=0$, so that $\bar\theta_\fb-\phi_c=0$ [see Eq.~\rp{theta}], which corresponds to the cavity amplitude quadrature which is directly coupled to the mechanical resonator.

\subsubsection{
Effectively reduced cavity linewidth}\label{effectivecavity}

In general the effective susceptibility defined in Eq.~\rp{chieff} exhibits many resonances due to the feedback term $\lambda_{c,\fb}(\omega)$ (see Sec.~\ref{squashing}). However, in this case the system response is constrained by the cavity linewidth $\kappa$, so that only the resonances which fall within the cavity linewidth are relevant.
As discussed in Secs.~\rp{squashing} and \rp{S_i_cav},
the spacing between these resonances depends upon the delay time.
In particular, if the delay time is sufficiently short for the distance between anti--squashing peaks to be much larger than the cavity linewidth, only a single resonance of the feedback system, which falls within the cavity bandwidth, is relevant.


Here
we want to identify the relevant resonance in the system response when the feedback delay time is small $\tau_\fb\ll{1}/{\kappa}$.
In general the effective susceptibility is given by
\begin{eqnarray}\label{chiceff}
\chi_c^{\rm eff}(\omega)&=&\lambda_{c,\fb}(\omega)\ \chi_c(\omega)
\\&&\hspace{-1cm}\,=
\pg{
\pq{\kappa+\ii(\Delta-\omega)}
-\mu_\fb(\omega)\pq{  \ee^{-\ii\bar\theta_\fb}+
\frac{\kappa+\ii(\Delta-\omega)}{\kappa-\ii(\Delta+\omega)}\ee^{\ii\bar\theta_\fb}
}
}^{-1}.
\nn
\end{eqnarray}
We look for the single pole of this function which characterizes the system dynamics close to the detuning frequency $\omega\sim\Delta$.
Hence we can define $\omega=\Delta+\delta$ and assume $\delta\ll\Delta$. Expanding $\chi_c^{\rm eff}(\omega)^{-1}$ at lowest order in $\delta$, 
we find that $\chi_c^{\rm eff}(\omega)$ can be approximated as
\begin{eqnarray}
\chi_c^{\rm eff}(\omega)&\simeq&\frac{
\ii
}{u\ \pt{\omega-\wt \nu}
}\ ,
\nn\\
\end{eqnarray}
where
\begin{eqnarray}
u=1+\frac{2\,\ii\,\Delta\,\ee^{\ii\,\bar\theta_\fb}}{\pt{\kappa-2\,\ii\,\Delta}^2}\ \mu_\fb(\Delta)-\ii\pt{
\ee^{-\ii\,\bar\theta_\fb}+
\frac{\kappa\,\ee^{\ii\,\bar\theta_\fb}}{\kappa-2\,\ii\,\Delta}
}\ \mu_\fb'(\Delta)\ ,
\end{eqnarray}
with
$\mu_\fb'(\omega)=\partial{\mu_\fb(\omega)}/{\partial\omega}$,
and where $\wt \nu$ is the complex pole defined as
\begin{eqnarray}
\wt \nu=\Delta- \frac{\ii}{u}\pq{
\kappa-\pt{
\ee^{-\ii\,\bar\theta_\fb}+
\frac{\kappa\,\ee^{\ii\,\bar\theta_\fb}}{\kappa-2\,\ii\,\Delta}
}\ \mu_\fb(\Delta)
}\ .
\end{eqnarray}
The effective system decay rate and detuning are therefore given by
\begin{eqnarray}\label{Deltaeff}
\Delta_{\rm eff}&=&{\rm Re}\pg{\wt\nu}\ ,
 \nn\\
\kappa_{\rm eff}&=&-{\rm Im}\pg{\wt\nu}\ ,
\end{eqnarray}
and
\begin{eqnarray}
\chi_c^{\rm eff}(\omega)\simeq\frac{1}{u\ \pq{\kappa_{\rm eff}+\ii\pt{\Delta_{\rm eff}-\omega}}}\ .
\end{eqnarray}

%
If we further assume, as in Refs.~\cite{Rossi,Kralj,Rossi2}, that $\kappa\ll\Delta$, $\abs{\mu_\fb(\Delta)}\ll\Delta$, and $\abs{\mu_\fb'(\Delta)}\ll1$ [this is, for example, the case of the filter function defined in Eq.~\rp{gfbflat} where $\mu_\fb'(\omega)=2\,\ii\,\tau_\fb\sqrt{\kappa_\fb\,\kappa_1\,\eta}\ { \wt h_\fb(\omega)^2}/{\wt g_\fb(\omega)}$], with sufficiently short delay time, 
then $u\simeq 1$ and
\begin{eqnarray}\label{Deltaeffkappaeff}
\Delta_\eff&\simeq&\Delta-{\rm Im}\pq{\ee^{-\ii\bar\theta_\fb}\ \mu_\fb(\Delta)}\ ,
\nn\\
\kappa_\eff&\simeq&\kappa-{\rm Re}\pq{\ee^{-\ii\bar\theta_\fb}\ \mu_\fb(\Delta)}\
\end{eqnarray}
(see also the Supplementary material of Ref.~\cite{Rossi}).

\section{Feedback--controlled light with an optomechanical system}\label{SecOM}

\begin{figure}[t!]
\includegraphics[width=8.5cm]{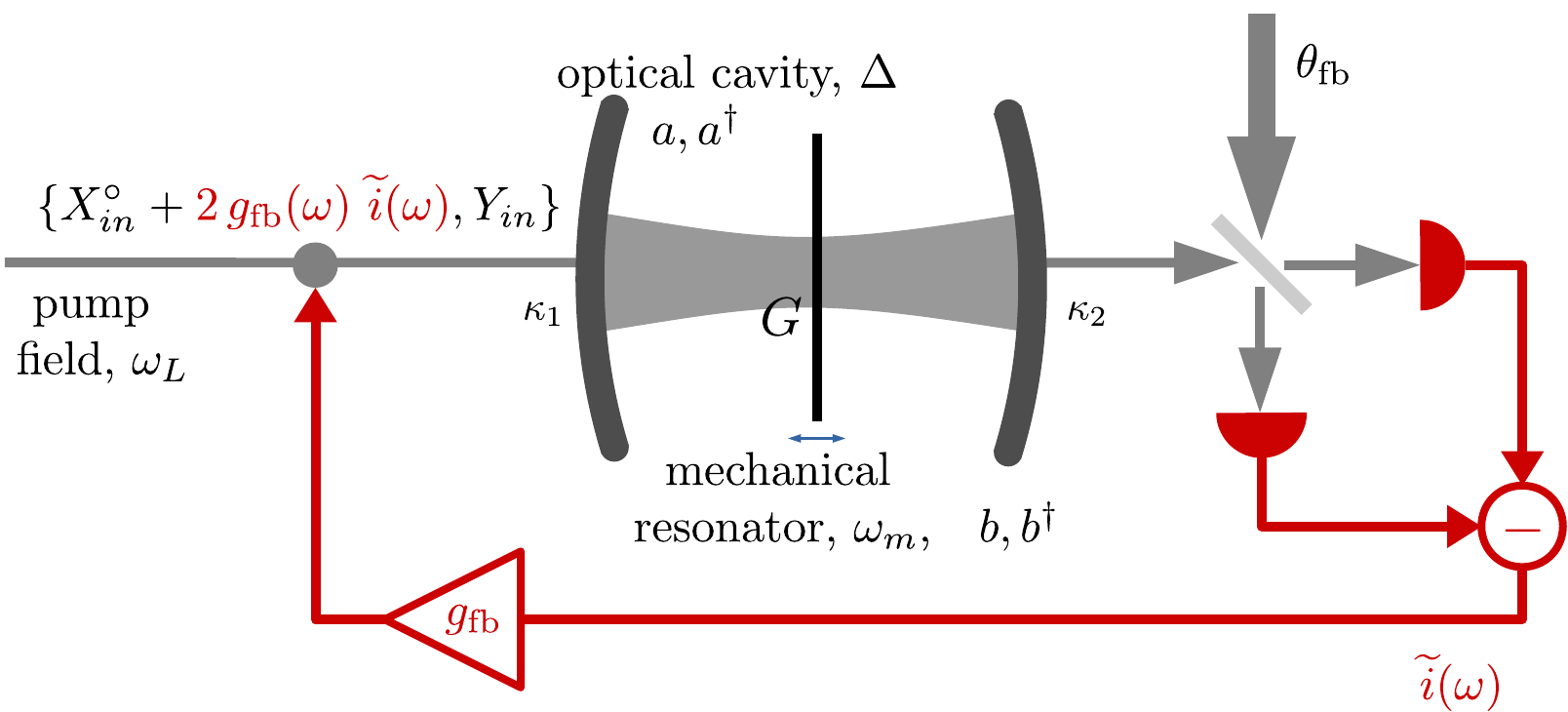}
\caption{The feedback loop: a quadrature at phase $\theta_\fb$ of the field transmitted through a cavity which contains a mechanical element (at frequency $\omega_{\rm m}$, and which interacts with the cavity field with strength $G$), is detected, and the corresponding photocurrent is used to modulate the input amplitude $X_\inn$. The feedback can be closed also by measuring the reflected field.
}\label{schema2}
\end{figure}

Let us now add a mechanical element within the optical cavity, as in Fig.~\ref{schema2}, and study the corresponding feedback--controlled optomechanical dynamics.

%



\subsection{The model}

The model of Sec.~\ref{modelcav} can be extended by including a vibrational mode, with frequency $\omega_{\rm m}$ and dissipation rate $\gamma\ll\omega_{\rm m}$, of a mechanical element which interacts by radiation pressure with the cavity light at strength $g_0$. In  particular, we consider the annihilation and creation operators $\wt b(\omega)$ and $\wt b\da(\omega)$ for the mechanical vibrations about the average position $\bar q=\sqrt{2}\,g_0\,\alpha_c^2/\omega_{\rm m}$ (relative to the mechanical position with no light), where $\alpha_c=\sqrt{2\,\kappa_1}\abs{\chi_c(0)}\alpha_\inn^\circ$ is the cavity field amplitude, with $\alpha_\inn^\circ$ the amplitude of the driving field and $\chi_c(0)$ the cavity susceptibility defined as in Eq.~\rp{chic}, but with the detuning $\Delta=\omega_c-\omega_L-\sqrt{2}\,g_0\,  \bar q$ which here includes also the light shift due to the optomechanical interaction. Thereby, the linearized equation for the cavity field, which includes only the linear terms in the field and mechanical variables, with linearized interaction strength $G=g_0\,\alpha_c$, is given  by
\begin{eqnarray}\label{aomega2}
&&\hspace{-0.3cm}
-\pq{\kappa+\ii\pt{\Delta-\omega}}\wt a(\omega)+\ii\,G\pq{\wt b\da(\omega)+\wt b(\omega)}+
\\&&\hspace{1cm}
+\sqrt{2\kappa_0}\,\wt a_{\rm in,1}(\omega)\,\ee^{-\ii\,\phi_c}+\sqrt{2\kappa_2}\,\wt a_{\rm in,2}(\omega)\,\,\ee^{-\ii\,\phi_c}\ =\ 0\ ,
\nn
\end{eqnarray}
where $\phi_c$ is the phase difference between the input and cavity field defined in Eq.~\rp{phic},
and the corresponding equation for the mechanical vibrations is
\begin{eqnarray}\label{bomega}
&&\hspace{-0.3cm}
-\pq{\frac{\gamma}{2}+\ii\pt{\omega_{\rm m}-\omega}}\wt b(\omega) +\ii\,G\pq{\wt a(\omega)+\wt a\da(\omega)}+
\\&&\hspace{5.5cm}
+\sqrt{\gamma}\,\wt b_\inn(\omega)\ =\ 0\ ,
\nn
\end{eqnarray}
where we have introduced the mechanical thermal noise operator $\wt b_\inn(\omega)$ characterized by the correlation functions $\av{\wt b_\inn(\omega)\ \wt b_\inn(\omega')}=0$ and $\av{\wt b_\inn(\omega)\ \wt b_\inn\da(\omega')}=(n_{\rm th}+1)\ \delta\pt{\omega+\omega'}$, with $n_{\rm th}$ the number of thermal excitations.

\subsubsection{Feedback photocurrent with an optomechanical system}

In general, the formula for the feedback photocurrent has the same structure as the one in Eqs.~\rp{gi} and \rp{lambda_fb} (which are valid for an empty cavity), that is
\begin{eqnarray}
\wt g_\fb(\omega)\ \wt i(\omega)=\wt h_\fb(\omega)\ \lambda_{\om,\fb}(\omega)\ \wt i^\circ\pt{\omega}\ ,
\end{eqnarray}
but with the squashing factor now being
\begin{eqnarray}
\lambda_{\om,\fb}(\omega)=\frac{1}{1-2\,\mu_\fb(\omega)\,\zeta_{\om}\al{\bar\theta_\fb}(\omega)}
\end{eqnarray}
where $\zeta_{\om}\al{\bar\theta_\fb}(\omega)$ is the cavity response function modified by the mechanical resonator [see Eq.~\rp{zetac} for the empty cavity case]. It is explicitly given by
\begin{eqnarray}\label{zetaom}
\zeta_{\om}\al{\bar\theta_\fb}(\omega)&=&2\,
\zeta_{\rm m}^{G}(\omega)
\lpq{
\zeta_c\al{\bar\theta_\fb}(\omega)\ \zeta_{\rm m}(\omega)^{-1}\
}\\&&\rpq{
+4\, G^2\,\cos(\phi_c)\,\sin(\bar\theta)\, \chi_c(\omega)\,\chi_c(-\omega)^*
}\ ,
\nn
\end{eqnarray}
where $\bar\theta_\fb$ is defined in Eqs.~\rp{theta}, $\bar\theta=\bar\theta_\fb-\phi_c$, $\zeta_{\rm m}(\omega)$ is the mechanical response function defined in terms of
$\chi_{\rm m}(\omega)=\pq{\frac{\gamma}{2}+\ii\pt{\omega_{\rm m}-\ii\,\omega}}^{-1}$ as
\begin{eqnarray}
\zeta_{\rm m}(\omega)&=&\ii\ \frac{\chi_{\rm m}(\omega)-\chi_{\rm m}(-\omega)^*}{2}
\end{eqnarray}
[such that when $\gamma\ll\omega_{\rm m}$, $\zeta_{\rm m}(\omega)
\simeq {\omega_{\rm m}}/(\omega_{\rm m}^2-\omega^2-\ii\,\omega\,\gamma)$], and where we have also introduced
the mechanical response function modified by the optomechanical interaction
\begin{eqnarray}\label{Xicirc}
\zeta_{\rm m}^{G}(\omega)=\pq{\zeta_{\rm m}(\omega)^{-1}-4\ G^2\ \zeta_c\al{-\pi/2}(\omega)}^{-1}\ ,
\end{eqnarray}
with the cavity response function $\zeta_c\al{-\pi/2}(\omega)$ defined in Eq.~\rp{zetac}.
We further note that in this case the power spectrum of the
photocurrent without feedback $\wt i^\circ(\omega)$ is not equal to one, but it is frequency dependent including also the effect of the mechanical thermal noise. Hence, if we define $S_{\out,\fb}^{\circ\,(\theta_\fb)}(\omega)$ as the power spectrum of the output field with no feedback 
such that
$
\av{
\wt X_{\out,\fb}^{\circ\,\pt{\theta_\fb}}(\omega)\ \wt X_{\out,\fb}^{\circ\,\pt{\theta_\fb}}(\omega')
}
=\delta\pt{\omega+\omega'}\ S_{\out,\fb}^{\circ\,(\theta_\fb)}(\omega)
$ (the specific form of which is cumbersome and not relevant here), then the power spectrum of the feedback photocurrent takes the form
 \begin{eqnarray}\label{Sifull}
S_i(\omega)=\abs{\lambda_{\om,\fb}(\omega)}^2\ \pg{1+\eta\,\pq{S_{\out,\fb}^{\circ\,(\theta_\fb)}(\omega)-1}}\ .
\end{eqnarray}

\subsubsection{Mechanical vibrations}

Eqs.~\rp{aomega2} and \rp{bomega} can be solved to determine the expression  for the mechanical position operator $\wt q(\omega)=\pq{\wt b(\omega)+\wt b\da(\omega)}/{\sqrt{2}}$ which can be written as
\begin{eqnarray}
\wt q(\omega)&=&
\zeta_{\rm m,\fb}^{G}(\omega)\
\pg{\wt\xi_{\rm m}(\omega)
+ \wt\xi_{c,\fb}(\omega)
}\ ,
\end{eqnarray}
where we have introduced the mechanical response function modified by both the optomechanical interaction and the feedback [see Eq.~\rp{Xicirc} for the equivalent equation without feedback]
\begin{eqnarray}\label{Xi}
\zeta_{\rm m,\fb}^{G}(\omega)=\pq{\zeta_{\rm m}(\omega)^{-1}-4\ G^2\ \zeta_{c,\fb}(\omega)}^{-1}\ ,
\end{eqnarray}
in which also the cavity response function now includes the effect of the feedback according to the relation
\begin{eqnarray}\label{zetacfb}
\zeta_{c,\fb}(\omega)&=&
\lambda_{c,\fb}(\omega)\,\zeta_c\al{-\pi/2}(\omega)
\\&&\times
\pq{1
+2\,\mu_\fb(\omega)\
\chi_c(\omega)\,\chi_c(-\omega)^*\ \cos(\phi_c)\ \sin(\bar\theta)
}\ .
\nn
\end{eqnarray}
Moreover we have also introduced the mechanical and electromagnetic noise terms $\wt\xi_{\rm m}(\omega)$ and $\wt\xi_{c,\fb}(\omega)$. In particular the first is defined as
\begin{eqnarray}
\wt\xi_{\rm m}(\omega)=\sqrt{\frac{\gamma}{2}}\ \frac{\chi_{\rm m}(\omega)\ \wt b_{\rm in}(\omega)+\chi_{\rm m}(-\omega)^*\ \wt b_{\rm in}\da(\omega)}{\zeta_{\rm m}(\omega)}
\end{eqnarray}
so that, when $\gamma\ll\omega_{\rm m}$, the corresponding symmetrized power spectrum can be approximated (in the relevant range of frequencies close to the mechanical frequency) as
$\int\,\dd\omega\pq{\av{\wt\xi_{\rm m}(\omega)\ \wt\xi_{\rm m}(\omega')}+\av{\wt\xi_{\rm m}(-\omega)\ \wt\xi_{\rm m}(\omega')} }/2\simeq \gamma\ (2\,n_T+1)\equiv S_{th}$. Instead, the electromagnetic noise $\wt\xi_{c,\fb}(\omega)$ includes also the noise introduced by the feedback process and can be written as
\begin{eqnarray}
\wt\xi_{c,\fb}(\omega)&=&2\, G\lpq{
\ee^{-\ii\,\phi_c}\ \chi_c^\eff(\omega)\ \wt f_{\rm in,c}(\omega)
}
\nn\\&&\hspace{2cm}\rpq{
+\ee^{\ii\,\phi_c}\ \chi_c^\eff(-\omega)^*\ \wt f_{\rm in,c}\da(\omega)
},
\end{eqnarray}
where the effective cavity susceptibility $\chi_c^\eff(\omega)$ and the total cavity noise operator $\wt f_{\rm in,c}(\omega)$ are introduced in Eqs.~\rp{chieff} and \rp{fcin} respectively.

\subsubsection{Power spectrum of the  mechanical position}

The corresponding mechanical power spectrum can be detected by measuring the phase modulation of an additional probe field resonant with the cavity mode.
%
%
Specifically, the power spectrum of the field is proportional to the symmetrized position spectrum of the  mechanical position,
which is given by
\begin{eqnarray}\label{Sq}
S_q(\omega)&=&\int\,\dd\omega'\frac{\av{\wt q(\omega)\ \wt q(\omega')} + \av{\wt q(-\omega)\ \wt q(\omega')}}{2}
\nn\\&&\hspace{-1.5cm}=
\abs{\zeta_{\rm m,\fb}^{G}(\omega)}^2
\pq{
S_{\rm th}+S\al{\fb,0
}_{\rm rp}(\omega)+S\al{\fb,I}_{\rm rp}(\omega)+ S\al{\fb,II}_{\rm rp}(\omega)
}
\end{eqnarray}
where $S_{\rm th}\simeq \gamma\,\pt{2\,n_{\rm th}+1}$ and the radiation pressure contribution is divided into three terms
\begin{eqnarray}
S_{\rm rp}\al{\fb,0
}(\omega)&=&2\,G^2\,\kappa\pq{\abs{\chi_{c,\eff}(\omega)}^2+\abs{\chi_{c,\eff}(-\omega)}^2},
\nn\\
S_{\rm rp}\al{\fb,I}(\omega)&=&\frac{4\,G^2\,\abs{\mu_\fb(\omega)}^2}{\kappa_\fb\ \eta}\ \abs{\lambda_{c,\fb}(\omega)\ \zeta_c\al{\phi_c}}^2
-4\,G^2\ \abs{\lambda_{c,\fb}(\omega)}^2
\nn\\&&\hspace{-1.5cm}\times\
\pq{
\mu_\fb(\omega)\ \zeta_c\al{\phi_c}(\omega)\ \zeta_c\al{-\bar\theta}(\omega)^*
+
\mu_\fb(\omega)^*\ \zeta_c\al{\phi_c}(\omega)^*\ \zeta_c\al{-\bar\theta}(\omega)
},
\nn\\
S\al{\fb,II}_{\rm rp}(\omega)&=& 16\,\kappa\,G^2\,\abs{\mu_\fb(\omega)\
\lambda_{c,\fb}(\omega)\ \chi_c(\omega)\ \chi_c(-\omega)^*}^2
\sin^2(\bar\theta)
\nn\\&&\hspace{-0.3cm}
+16\,\kappa\,G^2\ 
\abs{\lambda_{c,\fb}(\omega)}^2\
\nn\\&&\hspace{0cm}\times
{\rm Re}\pq{
\mu_\fb(\omega)\
\chi_c(\omega)\,\chi_c(-\omega)^*\ \zeta_c\al{\phi_c-\frac{\pi}{2}}(\omega)^*
}\ \sin(\bar\theta)
\nn\\&&\hspace{-0.3cm}
-16\,G^2\ \abs{\mu_\fb(\omega)\ \lambda_{c,\fb}(\omega)}^2\ 
\\&&\hspace{0cm}\times
{\rm Re}\pq{
\chi_c(\omega)\,\chi_c(-\omega)^*\ \zeta_c\al{\phi_c}(\omega)^*
}
\ \sin(\bar\theta_\fb)\ \sin(\bar\theta),
\nn
\end{eqnarray}
where $\mu_\fb(\omega)$ is the filter feedback function defined in Eq.~\rp{mufb}.
As discussed in Sec.~\ref{Sec:EffectiveCavity}, the feedback modifies the cavity susceptibility and adds additional cavity noise. In turn this is reflected in a modified mechanical susceptibility [see Eq.~\rp{Xi}], and a modified radiation pressure noise term in the position spectrum, corresponding to the last three terms in Eq.~\rp{Sq}. The first one, $S_{\rm rp}\al{\fb,0}(\omega)$, accounts for the effect of the modified cavity susceptibility $\chi_{c,\eff}(\omega)$ and reduces to the standard radiation pressure term in the limit of zero feedback gain [i.e $\chi_{c,\eff}(\omega)\to\chi_c(\omega)$ when $\mu_\fb(\omega)\to0$]. The other two are instead due to the additional cavity noise. Here they are presented as two separated terms because of the different dependence on the homodyne phase.  
Specifically,
%
the last term, $S_{\rm rp}\al{\fb,II}(\omega)$, is zero when $\bar\theta=\bar\theta_\fb-\phi_c=0$ [with $\bar\theta_\fb$ the phase of the detected quadrature, $\theta_\fb,$ plus the phase shift of the output field as defined in Eq.~\rp{theta}] as in the case studied in Ref.~\cite{Rossi2}.

Let us now assume that the feedback is operated close to the instability, with a single feedback peak within the cavity bandwidth ($\kappa\ll1/\tau_\fb$) such that it is possible to define an effective cavity, as discussed in Sec.~\ref{effectivecavity}, with $\kappa_\eff,\abs{\Delta-\Delta_\eff}\ll\kappa,\omega_{\rm m}$. Then we can approximate
$\chi_{c,\eff}(\omega)\ \chi_{c,\eff}(-\omega)\sim 0$,
while $\lambda_{c,\fb}(\omega)\ \chi_c(\omega)\ \chi_c(-\omega)^*$ has two peaks at $\omega\sim\pm\Delta$, such that
$\lambda_{c,\fb}(\omega)\ \chi_c(\omega)\ \chi_c(-\omega)^*\sim
\frac{\chi_{c,\eff}(\omega)}{\kappa-2\,\ii\,\Delta}+\frac{\chi_{c,\eff}(-\omega)^*}{\kappa+2\,\ii\,\Delta}$.
Using these approximations we find the following approximated spectra
\begin{eqnarray}
S_{\rm rp}\al{\fb,I}(\omega)&\simeq&2\,G^2\,\pq{Z_I(\omega)\,\abs{\chi_{c,\eff}(\omega)}^2+Z_I(-\omega)\,\abs{\chi_{c,\eff}(-\omega)}^2}
\nn\\
\end{eqnarray}
with
\begin{eqnarray}
Z_I(\omega)&=&\frac{\abs{\mu_\fb(\omega)}^2}{2\,\kappa_\fb\ \eta}-{\rm Re}\pg{
\mu_\fb(\omega)\ \ee^{-\ii\,\bar\theta_\fb}
}
\end{eqnarray}
(see also the supplemental material of Ref.~\cite{Rossi2}),
and
\begin{eqnarray}
S\al{\fb,II}_{\rm rp}(\omega)&=&2\,G^2\,\pq{Z_{II}(\omega)\,\abs{\chi_{c,\eff}(\omega)}^2+Z_{II}(-\omega)\,\abs{\chi_{c,\eff}(-\omega)}^2},
\nn\\
\end{eqnarray}
with
\begin{eqnarray}
Z_{II}(\omega)&=&\frac{8\,\kappa\,\abs{\mu_\fb(\omega)}^2\ \sin^2(\bar\theta)}{\kappa^2+4\,\Delta^2}
\nn\\&&
+4\,\kappa\ \sin(\bar\theta)\ {\rm Re}\pq{-\ii\,\frac{\mu_\fb(\omega)\ \ee^{\ii\,\phi_c}}{\kappa-2\,\ii\,\Delta}}
\nn\\&&
-4\ \abs{\mu_\fb(\omega)}^2\ \sin(\bar\theta_\fb)\ \sin(\bar\theta)\ {\rm Re}\pq{\frac{\ee^{\ii\,\phi_c}}{\kappa-2\,\ii\,\Delta}}\   .
\end{eqnarray}
Finally, assuming a sufficiently broad filter function $\mu_\fb(\omega)$ almost constant over the cavity bandwidth $\kappa$, we can approximate the position spectrum as
\begin{eqnarray}
S_q(\omega)&=&
\abs{\zeta_{\rm m,\fb}^{G}(\omega)}^2
\pq{
S_{\rm th}+S\al{Z}_{\rm rp}(\omega)
},
\end{eqnarray}
where the radiation pressure term takes the form
\begin{eqnarray}
S_{\rm rp}\al{Z}(\omega)&\simeq&2\,G^2\,Z\pq{\abs{\chi_{c,\eff}(\omega)}^2+\abs{\chi_{c,\eff}(-\omega)}^2},
\end{eqnarray}
with
\begin{eqnarray}
Z=\kappa+Z_I(\Delta)+Z_{II}(\Delta)\ .
\end{eqnarray}
In Ref.~\cite{Rossi2} we have studied in detail the case in which the feedback is closed in transmission with $\bar\theta=\bar\theta_\fb-\phi_c=0$ (so that $Z_{II}(\omega)=0$) and $\kappa\ll\Delta$, so that the effective parameters are equal to those defined in Eq.~\rp{Deltaeffkappaeff}.
In that case (see also the supplemental material of Ref.~\cite{Rossi2}) we have been able to integrate analytically the position spectrum using the results of Ref.~\cite{Genes08},
and find a simple expression for the steady state number of mechanical excitations (it is not possible to directly apply the results of Ref.~\cite{Genes08} when $\bar\theta\neq0$ because also the mechanical response function $\zeta_{\rm m,\fb}^{G}(\omega)$ is modified in that case).
In particular, in Ref.~\cite{Rossi2} we have shown that,
while the mechanical resonator can be cooled to lower temperature with the help of feedback--controlled light as a result of the reduced effective cavity linewidth that is observed when the feedback is operated close to the mechanical instability, the cooling efficiency is degraded when the effective decay rate $\kappa_\eff$ is so low that $\kappa_\eff<G$. In this case, in fact, the mechanical energy can not be efficiently dissipated by the cavity, and the system enters a regime of strong coupling in which energy is coherently exchanged between the optical cavity and the mechanical resonator, with the consequent observation of normal mode splitting in the mechanical response~\cite{Rossi2}.

In the next section we will investigate this regime of effective strong coupling, and in particular discuss the onset of coherent optomechanical energy exchange.
Afterwards, in Sec.~\ref{cooling}, we will focus on the regime of optimal cooling, $\kappa_\eff>G$, where we analyse in detail the validity of the perturbative approach, based on the evaluation of light scattering rates, which we have employed in Ref.~\cite{Rossi}.

\subsection{Feedback-mediated strong coupling and coherent optomechanical oscillations}

In Ref.~\cite{Rossi2} we have shown that the reduced effective cavity linewidth experienced by the system in the anti-squashing regime close to the feedback instability can be used to promote the system to the strong coupling regime.
This entails that, as shown below, coherent light--matter oscillations are observable when, for example, a light pulse is injected into the cavity.

Specifically, here we study the response of the system to a short light pulse, and we study how it is transferred to the mechanical resonator.
We consider the optomechanical model of Eqs.~\rp{aomega2} and \rp{bomega}, in the time domain, and
include an additional driving pulse with sufficiently small amplitude 
for the linearised description to still be valid. 
The pulse acts on the cavity field at time $t=0$ and is 
much shorter than the system dynamics timescale, so that can be described by
an input driving term of the form $\sqrt{2\,\pi}\,\alpha_p\ \delta(t)$.
%
The corresponding equation for the field amplitude $\alpha(t)=\av{a(t)}$ (which is zero in the previous cases)
is then given by
\begin{eqnarray}
\dot\alpha(t)&=&-\pt{\kappa+\ii\Delta}\alpha(t)+\ii\, G\ \pq{\beta(t)+\beta(t)^*}
\nn\\&&
+\sqrt{2\kappa_1}\pq{\sqrt{2\,\pi}\,\alpha_p\,\delta(t)+\av{\Phi(t)}}\ \ee^{-\ii\phi_c}
\end{eqnarray}
with the boundary condition $\alpha(t)=0$ for $t\leq 0$, $\beta(t)=\av{b(t)}$, and where the feedback term $\Phi(t)$ is introduced in Eq.~\rp{Phi}. Assuming a flat feedback filter function as the one defined in Eq.~\rp{gfbflat} with $\phi_\fb=0$, so that $g_\fb(t)=2\sqrt{2\pi}\ \bar g_\fb\ \delta(t-\tau_\fb)$,
we have
\begin{eqnarray}
\av{\Phi(t)}&=&
\sqrt{2\ \kappa_\fb\,\eta}\ \bar g_\fb\ \pq{\alpha(t-\tau_{\rm fb})\,\ee^{-\ii\,\bar\theta}+\alpha(t-\tau_{\rm fb})^*\,\ee^{\ii\,\bar\theta}}\ .
\nn\\
\end{eqnarray}
Moreover, the mechanical variable $\beta(t)=\av{b(t)}$ fulfils the equation
\begin{eqnarray}
\dot\beta(t)=-\pt{\frac{\gamma}{2}+\ii\omega_{\rm m}}\beta(t)+\ii\,G\,\pq{\alpha(t)+\alpha(t)^*}\ ,
\end{eqnarray}
with the boundary condition $\beta(t)=0$ for $t\leq 0$.
We further assume that the cavity is close to resonance with the red sideband transition $\Delta\sim\omega_{\rm m}$, and we decompose the system variables
as $\alpha(t)=\bar\alpha(t)\ \ee^{-\ii\,\omega_{\rm m}\,t}$ and $\beta(t)=\bar\beta(t)\ \ee^{-\ii\,\omega_{\rm m}\,t}$ where $\bar\alpha(t)$ and $\bar\beta(t)$ are slowly varying amplitudes.
As a consequence we have the coupled equations
\begin{eqnarray}
\dot{\bar\alpha}(t)&=&-\pt{\kappa+\ii\delta}\bar\alpha(t)+\ii\, G\ \pq{\bar\beta(t)+\bar\beta(t)^*\,\ee^{2\,\ii\,\omega_{\rm m}\,t}}
\nn\\&&
+2\,\sqrt{\kappa_1\,\pi}\,\alpha_p\,\delta(t)\ \ee^{-\ii\phi_c}
+\bar\mu_\fb\ \lpq{
\bar\alpha(t-\tau_{\rm fb})\ \ee^{-\ii\bar\theta_\fb}\,\ee^{\ii\,\omega_{\rm m}\,\tau_\fb}
}\nn\\&&\hspace{0cm}\rpq{
+
\bar\alpha(t-\tau_{\rm fb})^*\ \ee^{-\ii\pt{\bar\theta_\fb-2\,\phi_c}}\,\ee^{2\,\ii\,\omega_{\rm m}\,t}\ee^{-\ii\,\omega_{\rm m}\,\tau_\fb}
}\ ,
\nn\\
\dot{\bar\beta}(t)&=&-\frac{\gamma}{2}\,\bar\beta(t)+\ii\,G\,\pq{\bar\alpha(t)+\bar\alpha(t)^*\,\ee^{2\,\ii\,\omega_{\rm m}\,t}}\ ,
\end{eqnarray}
with
$\delta=\Delta-\omega_{\rm m}$ (that is of the same order or smaller than $\kappa$) and
$\bar\mu_\fb=2\,\sqrt{\kappa_\fb\,\kappa_1\,\eta}\ \bar g_\fb$.
We first note that when $\tau_\fb$ is sufficiently small for a single feedback peak to fall within the cavity linewidth, i.e. when $\tau_\fb\ll1/\kappa$, we can approximate the slowly varying cavity amplitude at time $t-\tau_\fb$ with that at time $t$, i.e. $\bar\alpha(t-\tau_\fb)\sim\bar\alpha(t)$, so that we can introduce the effective cavity parameters
\begin{eqnarray}
\kappa_\eff&=&\kappa-\bar\mu_\fb\ \cos(\omega_{\rm m}\ \tau_\fb-\bar\theta_\fb)\ ,
\nn\\
\delta_\eff&=&\delta-\bar\mu_\fb\ \sin(\omega_{\rm m}\ \tau_\fb-\bar\theta_\fb)
\end{eqnarray}
which are equivalent to those defined in Eq.~\rp{Deltaeffkappaeff}.
Moreover, we consider
the limit of large mechanical frequency $\omega_{\rm m}\gg G,\bar\mu_\fb$,
so that
we can neglect the non-resonant terms, and eventually we find
\begin{eqnarray}\label{eqalphabetaapprox}
\dot{\bar\alpha}(t)&=&-\pt{\kappa_\eff+\ii\delta_\eff}\bar\alpha(t)+\ii\, G\ \bar\beta(t)
+2\,\sqrt{\kappa_1\,\pi}\,\alpha_p\,\delta(t)\ \ee^{-\ii\phi_c}
\nn\\
\dot{\bar\beta}(t)&=&-\frac{\gamma}{2}\,\bar\beta(t)+\ii\,G\,\bar\alpha(t)\ .
\end{eqnarray}
These equations can be easily solved. In particular, when $\kappa_\eff\ll G$, they describe coherent oscillations between the optical cavity and the mechanical resonator (even if the original cavity linewidth is large $\kappa\gg G$),
according to the equations
\begin{eqnarray}\label{betaalphaoscillations}
\bar\beta(t)&\simeq&
\frac{\ii\,\sqrt{2\kappa_1}\,\alpha_p\,\ee^{-\ii\,\phi_c}}{
2
}\ \sin\pt{G\,t} \ee^{-\frac{\kappa_{\rm eff}}{2}\, t}
\label{betasinG}
\nn\\
\bar\alpha(t)
&\simeq&\frac{\sqrt{2\kappa_1}\,\alpha_p\,\ee^{-\ii\,\phi_c}}{
2}\ \cos\pt{G\,t}
\ \ee^{-\frac{\kappa_{\rm eff}}{2}\,t} \ ,
\end{eqnarray}
which are valid for $\delta_{\rm eff}\sim 0$ and for times much smaller than  $1/\gamma$.
They describe how
the initial optical amplitude is transferred to the mechanical resonator and then swapped back to the cavity, until it is eventually dissipated by cavity decay at rate $\kappa_{\rm eff}/2$.

\subsection{Sideband-Cooling}\label{cooling}

One of the central achievements of quantum optomechanics is the ability to cool a massive object to the quantum ground state of motion. In Refs.~\cite{Rossi,Kralj} we have shown that feedback--controlled light can significantly enhance the performance of sideband-cooling.
In particular, in Refs.~\cite{Rossi,Kralj} we have presented results based on the calculation of the Stokes and anti-Stokes scattering rates in terms of the spectrum of the cavity field fluctuations.
The scattering rates can then be used to determine
the cooling dynamics, which is valid in the weak coupling limit.
Specifically, in this limit the cavity acts as a noise source~\cite{Bowen}
with corresponding noise operator $F(t)$, given by the amplitude quadrature without the mechanical resonator
\begin{flalign}
F(t)\equiv \pq{a(t)+a\da(t)}\Bigl|_{G=0}\ .
\end{flalign}
Its power spectrum
\begin{eqnarray}\label{SF}
S_F(\omega)=\int_{-\infty}^\infty\dd\tau\ \ee^{-\ii\omega \tau}\av{F(0)\ F(\tau)}_{\rm st}
\end{eqnarray}
(where the label ``st'' indicates that the average is performed over the steady state) determines the rates $A_\pm=G^2\ S_{F}(\mp\omega_{\rm m})$ at which
mechanical excitations are transferred from the noise source (the cavity) to the resonator and  the other way round respectively. Thereby, the population of the mechanical state with $n$ excitations $p_n$ follows the standard rate equation $\dot p_n=-\pq{n\,\bar A_-+(n+1)\bar A_+}p_n+(n+1)\bar A_-\,p_{n+1}+n\,\bar A_+\,p_{n-1}$, with $\bar A_+=A_++\gamma\,n_{\rm th}$ and $\bar A_-=A_-+\gamma(n_{\rm th}+1)$, which implies
that the equation for the number of mechanical excitations $n(t)=\av{b\da(t)\ b(t)}$ is
\begin{eqnarray}\label{SecIII:dotn}
\dot n(t)=-\pt{\Gamma+\gamma}n(t)+A_++\gamma\,n_{\rm th}\ ,
\end{eqnarray}
with $\Gamma=A_--A_+$.
The corresponding
steady state number of mechanical excitations is finally given by
\begin{eqnarray}\label{nm}
n_{\rm m}=\frac{\gamma\ n_{\rm th}+\Gamma\ n_o}{\gamma+\Gamma},
\end{eqnarray}
where $n_o=\frac{A_+}{A_--A_+}$ define the backaction limit.
This is a general approach that has been successfully used to describe the cooling dynamics of mechanical resonators in various situations.
It is easy to show, by a standard adiabatic elimination of the cavity field,  that this approach is valid also with feedback--controlled cavities.
In particular, one can
consider the optomechanical model introduced in Eqs.~\rp{aomega2} and \rp{bomega}, and express the cavity variables in terms of the effective model for the cavity field defined in Eq.~\rp{Model2}.
When the cavity dynamics is fast as compared to the mechanical one, it is possible to eliminate the cavity degrees of freedom and obtain an equation for the mechanical resonator alone of the form
\begin{eqnarray}\label{SecIII:dotbarb}
\dot{b}(t)&=&-\pq{\frac{\gamma+\Gamma}{2}+\ii\pt{\omega_{\rm m}+\delta}}b(t)
+B_\inn(t)
+\sqrt{\gamma}\ b_\inn(t)\ ,
\nn\\
\end{eqnarray}
which includes the correction to the decay rate $\Gamma$, the frequency shift $\delta$ and the additional noise operator $B_\inn(t)$. The new parameters are corrections proportional to $G^2$ to the natural parameters of the resonator, which can be expressed in terms of the drift matrix of the effective cavity model [see Eq.~\rp{Model2}]
\begin{eqnarray}
\wt\MM(\omega)=\pt{\mat{cc}{
\wt\kappa_\eff(\omega)+\ii\,\wt\Delta_\eff(\omega) & \mu_\fb(\omega)\ \ee^{\ii\,\pt{\bar\theta_\fb-2\,\phi_c}} \\
\mu_\fb(\omega)\ \ee^{-\ii\,\pt{\bar\theta_\fb-2\,\phi_c}} & \wt\kappa_\eff(-\omega)-\ii\,\wt\Delta_\eff(-\omega)
}}
\end{eqnarray}
as $\Gamma=2\,G^2\ {\rm Re}\pg{ \pt{1,1}\pq{\wt\MM(\omega_{\rm m})-\ii\omega_{\rm m}\,\id}^{-1}\pt{1,-1}^T}$, and $\delta=G^2\ {\rm Im}\pg{ \pt{1,1}\pq{\wt\MM(\omega_{\rm m})-\ii\omega_{\rm m}\,\id}^{-1}\pt{1,-1}^T}$.
Moreover, the correlation functions of the additional noise operator $B_\inn(t)$ can be approximated, in the limit $\kappa\gg\Gamma+\gamma$, as $\av{B_{in}(t)\da\ B_{in}(t')}=\delta(t-t')\ G^2\ S_F(-\omega_{\rm m})$ and $\av{B_{in}(t)\ B_{in}(t')\da}=\delta(t-t')\ G^2\ S_F(\omega_{\rm m})$. This implies that the additional dissipation rate can also be expressed as $\Gamma=G^2\pq{ S_F(\omega_{\rm m})-S_F(\omega_{\rm m}) }$~\cite{Bowen}, and, in turn, this implies the validity of Eq.~\rp{SecIII:dotn}.

\subsubsection{Enhanced sideband cooling}\label{Sec.cooling}
\begin{figure}[t!]
\includegraphics[width=8.5cm]{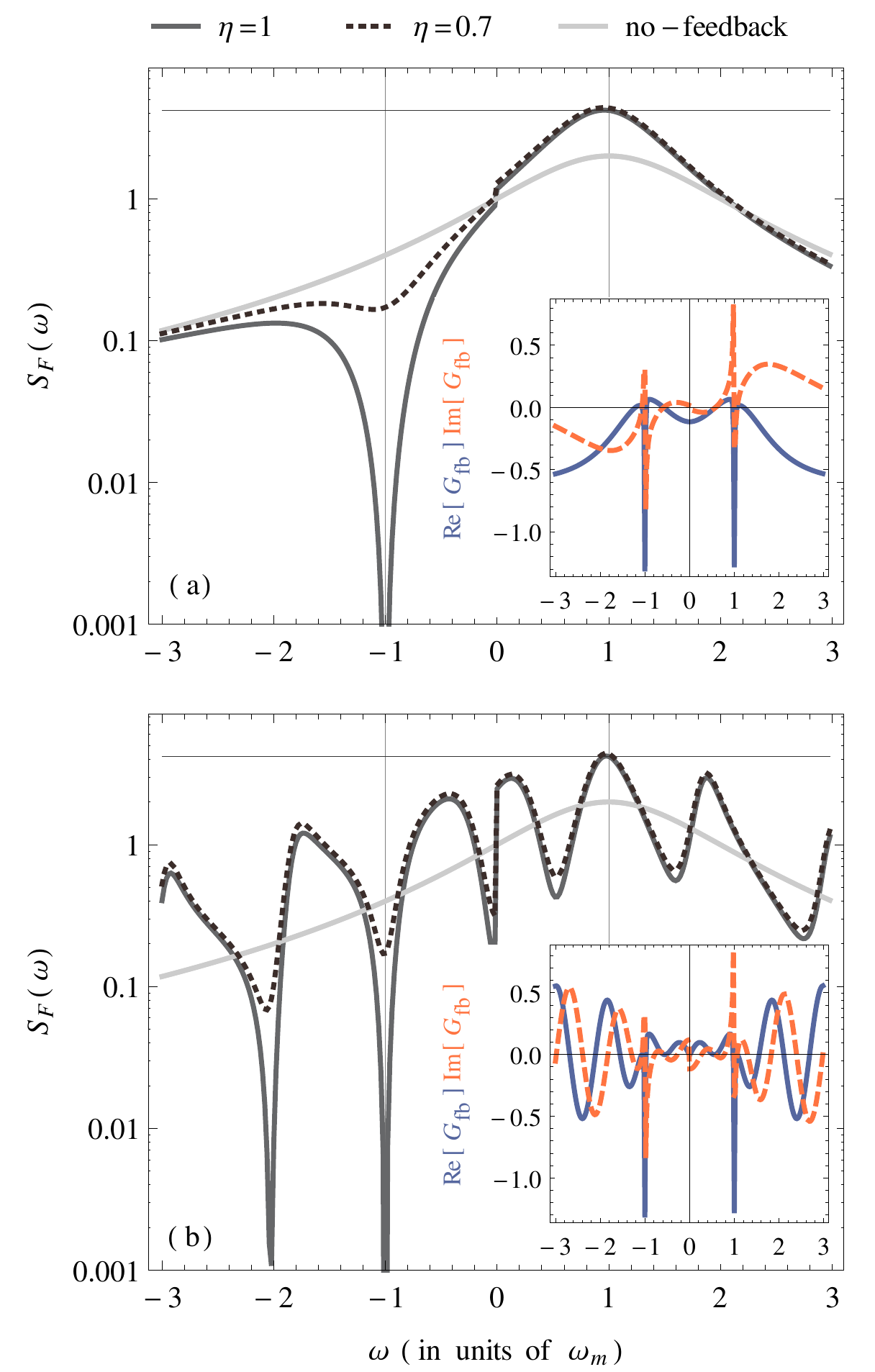}
\caption{Power spectrum of the cavity quadrature $F$ evaluated for a single--sided cavity ($\kappa_\fb=\kappa$ and feedback closed in reflection).
These plots are evaluated for the flat feedback filter function defined in Eq.~\rp{gfbflat}, and for the feedback parameters ($\bar g_\fb$ and $\phi_\fb$) that suppress anti-Stokes scattering according to Eq.~\rp{antiStokesSuppression} and for the value of $\theta_\fb$ that maximizes the corresponding Stokes scattering according to Eq.~\rp{optthetai}.
The corresponding total feedback transfer function $G_\fb(\omega)$ is reported in the insets (the blue solid lines correspond to the real part and the red dashed ones to the imaginary part), and they show that the feedback is stable (namely ${\rm Re}\pg{G_\fb}<1$ at the frequencies at which ${\rm Im}\pg{G_\fb}=0$).
Plot (a) refers to a short feedback delay time $\tau_\fb=0.1/\omega_{\rm m}$, and plot (b) to a  larger value $\tau_\fb=5/\omega_{\rm m}$.
The values of the curves at the frequencies $\pm\omega_{\rm m}$, indicated by the vertical lines, determine the values of the Stokes and anti-Stokes scattering rates, and they are equal for both cases (this indicates that the optimized values of the scattering rates are independent of the specific delay time).
The solid dark lines correspond to perfect detection efficiency $\eta=1$, and the dashed lines to $\eta=0.7$.
The light solid lines are the corresponding result without feedback (namely the result valid for standard sideband cooling).
The other parameters are $G=0.1\omega_{\rm m}$, $\Delta=\omega_{\rm m}$, and $\kappa=\omega_{\rm m}$.
}\label{fig.SF}
\end{figure}
\begin{figure*}[t!]
\includegraphics[width=18cm]{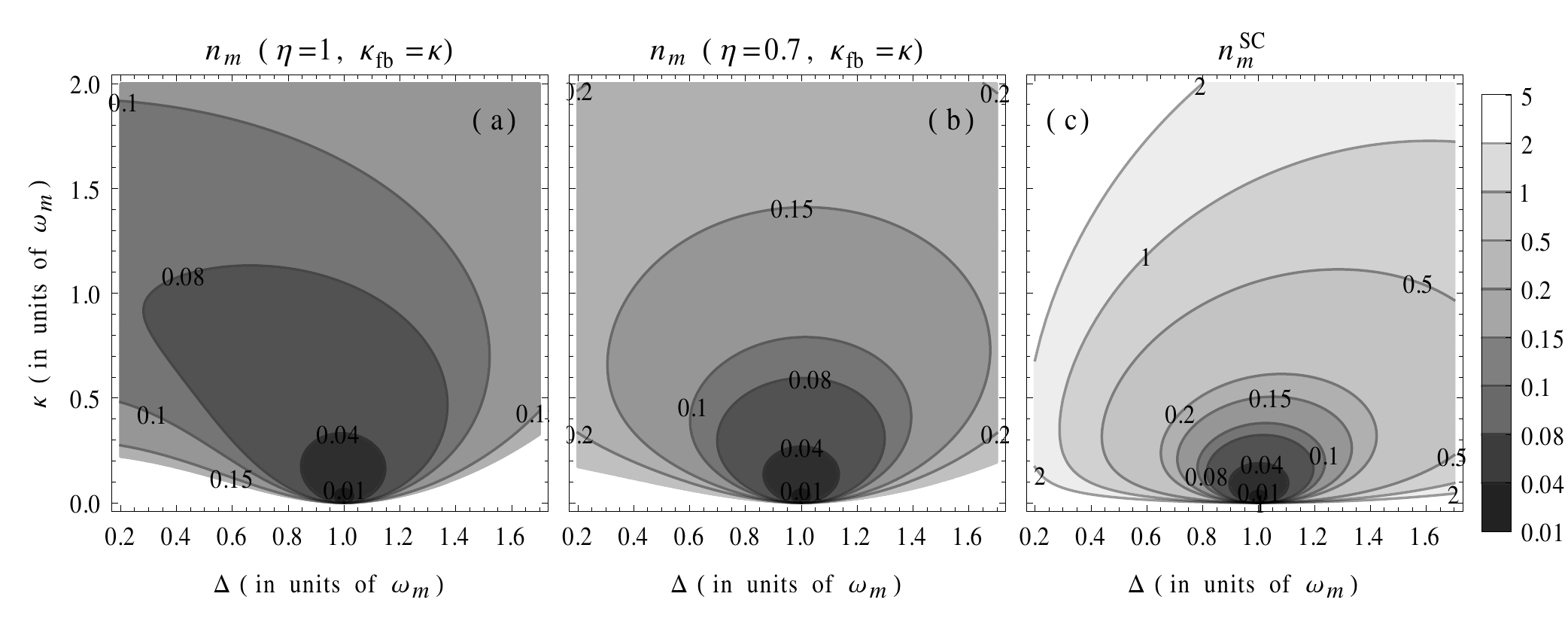}
\caption{(a), (b) Steady--state number of mechanical excitations as a function of the cavity detuning $\Delta$ and the cavity decay rate $\kappa$ evaluated (using the perturbative result of Eq.~\rp{nm} with the rates defined in Eq.~\rp{Apm}),  for a single-sided cavity (feedback closed in reflection), with the flat feedback filter function~\rp{gfbflat}, and with (a) perfect detection efficiency $\eta=1$, and (b) $\eta=0.7$. (c) Corresponding result for standard sideband cooling.
The feedback parameters ($\theta_\fb$, $\bar g_\fb$ and $\phi_\fb$)
are optimized at each point in order to achieve the optimal result of Eq.~\rp{optnm}. The other parameters are $\omega_{\rm m}=10$MHz, $\gamma=10^{-4}\omega_{\rm m}$, $n_{\rm th}=131$ (corresponding to a temperature of $100$mK) and $G=0.2\omega_{\rm m}$.
}\label{nst}
\end{figure*}

Explicit expressions for the Stokes and anti-Stokes rates, $A_\pm$,
can
be evaluated in terms of the power spectral matrix of the cavity field defined in Eq.~\rp{SSSva} as $A_\pm=G^2\ S_F(\mp\omega_{\rm m})=G^2\ (1,1) \SSS_\va(\mp\omega_{\rm m}) (1,1)^T$, and they are explicitly given by
\begin{eqnarray}\label{Apm}
A_\pm&=&
2\,G^2\ \kappa\ \lpq{
\abs{\chi_c(\mp\omega_{\rm m})
+\Lambda(\pm\omega_{\rm m})\
\ee^{\ii\pt{\bar\theta_\fb-\phi_c}}
}^2
}\\&&\rpq{\hspace{2cm}
+\pt{\frac{\kappa}{\kappa_\fb\ \eta}-1}\
\abs{\Lambda(\omega_{\rm m})}^2
}\ ,
\nn
\end{eqnarray}
where the effect of feedback is described by the coefficient
\begin{eqnarray}
\Lambda(\omega)=
\frac{\mu_\fb(\omega)}{\kappa}\
\lambda_{c,\fb}(\omega)\
\zeta_c\al{\phi_c}(\omega)
\end{eqnarray}
which fulfils the relation $\Lambda(\omega)^*=\Lambda(-\omega)$.

In Ref.~\cite{Rossi} we have identified two strategies to enhance sideband cooling which work in two distinct parameter regimes.

First, when thermal noise is 
low enough for the cooling efficiency to be limited 
by backaction noise [namely when the first term in the numerator of the equation for the steady state number of mechanical excitations~\rp{nm} is small, $\gamma\,n_{\rm th}\ll A_+$], it is convenient to suppress the rate for anti-Stokes scattering processes $A_+$.
This is achieved exploiting the destructive interference effect discussed in Sec.~\ref{Sec.inloop} which allows for the suppression of in--loop field fluctuations.
Specifically, the suppression is perfect in the limit of perfect detection efficiency ($\eta\sim1$) and when there is only one dissipation channel ($\kappa_\fb\sim\kappa$), such that the second term in the expression for the scattering rates~\rp{Apm} is negligible.
In fact, under this condition, when the feedback is properly selected, so that
\begin{eqnarray}\label{antiStokesSuppression}
\Lambda(\omega_{\rm m})\
\ee^{\ii\pt{\bar\theta_\fb-\phi_c}}=-\chi_c(-\omega_{\rm m})\ ,
\end{eqnarray}
then $A_+=0$.
Correspondingly,
using the fact that $\Lambda(-\omega)^*=\Lambda(\omega)$, we find
$A_-=2\,G^2\,\kappa\abs{\chi_c(\omega_{\rm m})-\chi_c(-\omega_{\rm m})^*\ \ee^{2\,\ii(\bar\theta_\fb-\phi_c)}}^2$.
In turn, the maximum of $A_-$ as a function of $\bar\theta_\fb$ is found for
\begin{eqnarray}\label{optthetai}
\ee^{2\,\ii\,\bar\theta_\fb}=
-\,\ee^{2\,\ii\,\phi_c}\
\frac{\kappa-\ii(\Delta-\omega_{\rm m})}{\sqrt{\kappa^2+(\Delta-\omega_{\rm m})^2}}\
\frac{\kappa-\ii(\Delta+\omega_{\rm m})}{\sqrt{\kappa^2+(\Delta+\omega_{\rm m})^2}}\ ,
\end{eqnarray}
such that $A_-$ reduces to
$A_-=2\,G^2\,\kappa\pq{\abs{\chi_c(\omega_{\rm m})}+\abs{\chi_c(-\omega_{\rm m})}}^2$.
%
In the general case of not perfect detection efficiency and multiple decay channels ($\eta<1$ and $\kappa_\fb<\kappa$), 
when the feedback is set to suppress the first term in the expression for $A_+$ in Eq.~\rp{Apm} 
[namely when Eq.~\rp{antiStokesSuppression} is fulfilled], 
we find $A_+=2\,G^2\,\kappa\pt{\frac{\kappa}{\eta\,\kappa_\fb}-1}\abs{\chi_c(\omega_{\rm m})}^2$, 
and the corresponding optimal value of the Stokes rate [determined by the condition in Eq.~\rp{optthetai}] is $A_-=2\,G^2\,\kappa\pq{\abs{\chi_c(\omega_{\rm m})}+\abs{\chi_c(-\omega_{\rm m})}}^2+A_+$.

Examples of the spectrum of fluctuations of the cavity field operator $S_F(\omega)$, which determines the values of the scattering rates $A_\pm$, are reported in Fig.~\ref{fig.SF} when the condition of anti-Stokes scattering suppression~\rp{antiStokesSuppression}  and the corresponding condition of optimal Stokes scattering~\rp{optthetai} are satisfied. They are reported for two values of the delay time [a short one in  Fig.~\ref{fig.SF} (a) and a longer one in (b)] and show that the scattering rates can be properly optimized to the same optimal values independently of the specific value of the delay times, namely in both plots the values $S_F(\pm\omega_{\rm m})$ are the same.

We further note that when Eqs.~\rp{antiStokesSuppression} and \rp{optthetai} are fulfilled,
the rates $A_\pm$ can be expressed in terms of the scattering rates with no feedback
$A_\pm^\circ=2\,G^2\kappa\abs{\chi_c(\mp\omega_{\rm m})}^2$ (namely the rates valid for standard sideband cooling) as
\begin{eqnarray}
A_+&=&\pt{\frac{\kappa}{\eta\,\kappa_\fb}-1}\ A_+^\circ,
\nn\\
A_-&=&\pq{\sqrt{A_-^\circ}+\sqrt{A_+^\circ}}^2+A_+\ .
\end{eqnarray}
Thereby, the corresponding steady state number of mechanical excitations is given by
\begin{eqnarray}\label{optnm}
n_{\rm m}=\frac{\gamma\ n_{\rm th}+\pt{\frac{\kappa}{\eta\,\kappa_\fb}-1}\ A_+^\circ}{\gamma+\pq{\sqrt{A_-^\circ}+\sqrt{A_+^\circ}}^2}\ .
\end{eqnarray}
This result shows that, when ${\kappa}/{\eta\kappa_\fb}-1<1$, the cooling efficiency can be significantly enhanced with respect to the standard sideband cooling result $n_{\rm  m}^{\rm SC}=\pt{\gamma\,n_{\rm th}+A_+^\circ}/\pt{\gamma+A_-^\circ-A_+^\circ}$; the enhancement is especially pronounced when the system is not in the resolved sideband regime.
Two examples of this result are reported in Figs.~\ref{nst} (a), for perfect detection efficiency, and (b), for reduced detection efficiency ($\eta=0.7$), and they are compared to the corresponding result of standard sideband cooling reported in plot (c).

In the opposite limit, when the back action noise is negligible [that is when the second term
in the numerator of the equation for the steady state number of mechanical excitations~\rp{nm} is negligible], the optimal cooling strategy
is to increase the value of the light--mediated mechanical dissipation rate $\Gamma=A_--A_+$. This can be achieved by operating the feedback close to instability, where both $A_-$ and $A_+$ are strongly enhanced. This is the limit that has been investigated also experimentally in Refs.~\cite{Rossi,Kralj}.
In particular, in the limit in which the cavity dynamics can be described by the effective susceptibility with the  effective parameters introduced in Eq.~\rp{Deltaeffkappaeff}, and the feedback parameters are properly set in order to achieve $\kappa_\eff\ll\kappa$ and $\Delta_\eff\simeq\Delta$, then the coefficient $\Lambda(\omega)$, which enters into the expressions for the scattering rates~\rp{Apm}, can be approximated as
\begin{eqnarray}
\Lambda(\omega_{\rm m})\simeq\frac{1}{2\ \kappa}\ \frac{\kappa-\kappa_\eff}{\kappa_\eff+\ii\pt{\Delta_\eff-\omega_{\rm m}}}\ \ee^{\ii\,\pt{\bar\theta_\fb-\phi_c}}\ .
\end{eqnarray}
Hence,
the Stokes and anti-Stokes scattering rates, for $\Delta_\eff=\omega_{\rm m}$, take the form
\begin{eqnarray}
A_+&\simeq&\frac{G^2\ \pt{\kappa-\kappa_\eff}^2}{2\,\eta\ \kappa_\fb\ \kappa_\eff^2}
\nn\\
A_-&\simeq& \frac{2\ G^2}{\kappa_\eff}+A_+\ .
\end{eqnarray}
Correspondingly, the steady state number of mechanical excitations,
expressed in terms of the standard sideband cooling result $n_{\rm m}^{\rm SC}=\gamma\ n_{\rm th}\ \kappa/(2\,G^2)$, is given by
\begin{eqnarray}
n_{\rm m}=n_{\rm m}^{\rm SC}\, \frac{\kappa_\eff}{\kappa}+\frac{\pt{\kappa-\kappa_\eff}^2}{4\,\eta\ \kappa_\fb\ \kappa_\eff}\ .
\end{eqnarray}
It reaches its minimum at $\kappa_\eff=\kappa/\sqrt{1+4\,\eta\,\kappa_\fb\ n_{\rm m}^{\rm SC}/\kappa}$, with the corresponding minimum value being
\begin{eqnarray}
n_{\rm m}=\frac{2\,n_{\rm m}^{\rm SC}}{1+\sqrt{1+4\,\eta\,\kappa_\fb\ n_{\rm m}^{\rm SC}/\kappa}}\ ,
\end{eqnarray}
which is strictly smaller than $n_{\rm m}^{\rm SC}$ (see also the Supplementary material of Ref.~\cite{Rossi}). 
Therefore, feedback \emph{always} allows to improve sideband cooling even in the regime dominated by thermal noise.


\subsection{Ponderomotive squeezing}

\begin{figure}[t!]
\includegraphics[width=8.5cm]{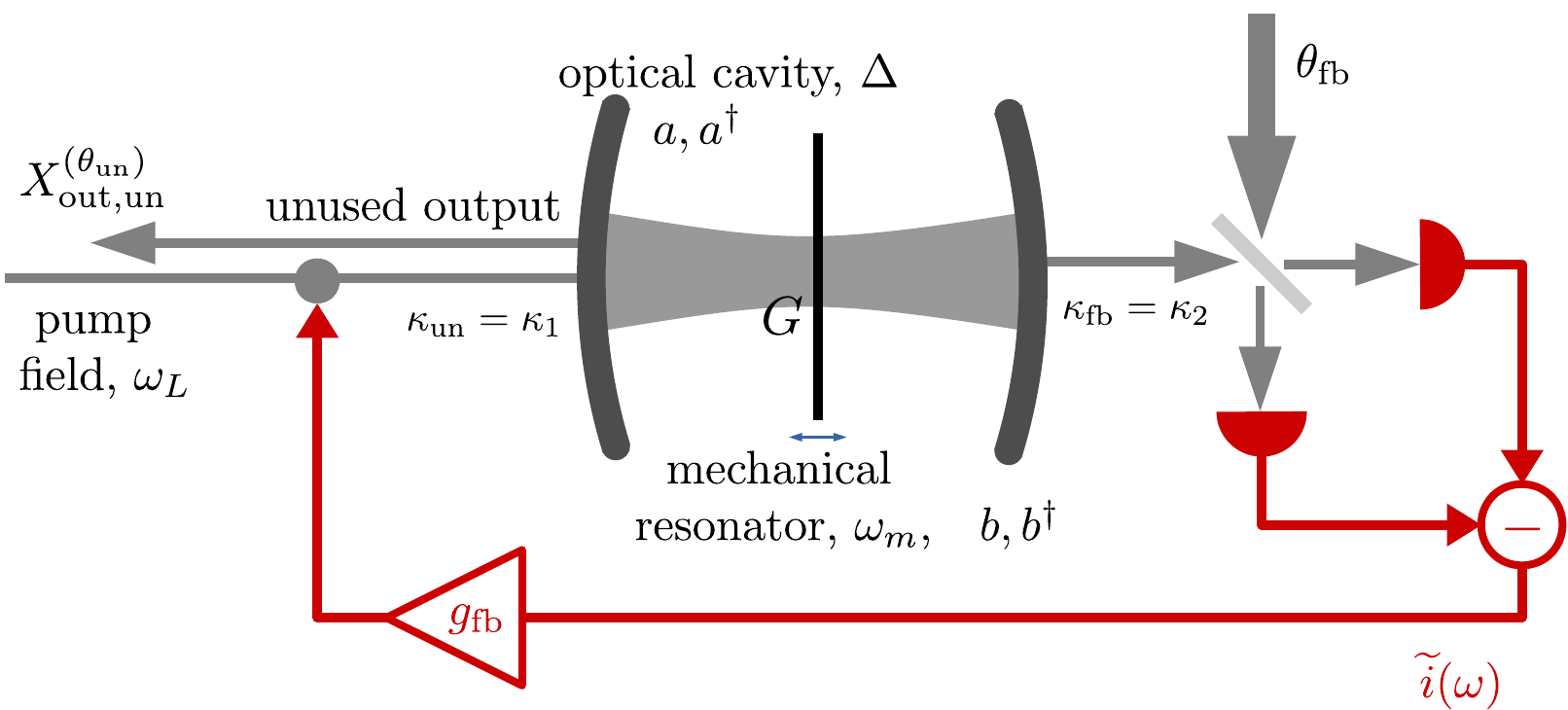}
\caption{Ponderomotive squeezing with feedback--controlled light: the squeezing of a quadrature of the reflected field at phase $\theta_\un$ is enhanced when the quadrature of the transmitted field at phase $\theta_\fb$ is detected, and the corresponding photocurrent is used to modulate the input amplitude $X_\inn$. The feedback can be also closed in reflection and in this case one would enhance the ponderomotive squeezing of the transmitted field, and the role of $\kappa_\fb$ and $\kappa_\un$ would be exchanged (i.e. $\kappa_\un=\kappa_2$ and $\kappa_\fb=\kappa_1$).
}\label{schema3}
\end{figure}

\begin{figure}[t!]
\includegraphics[width=8cm]{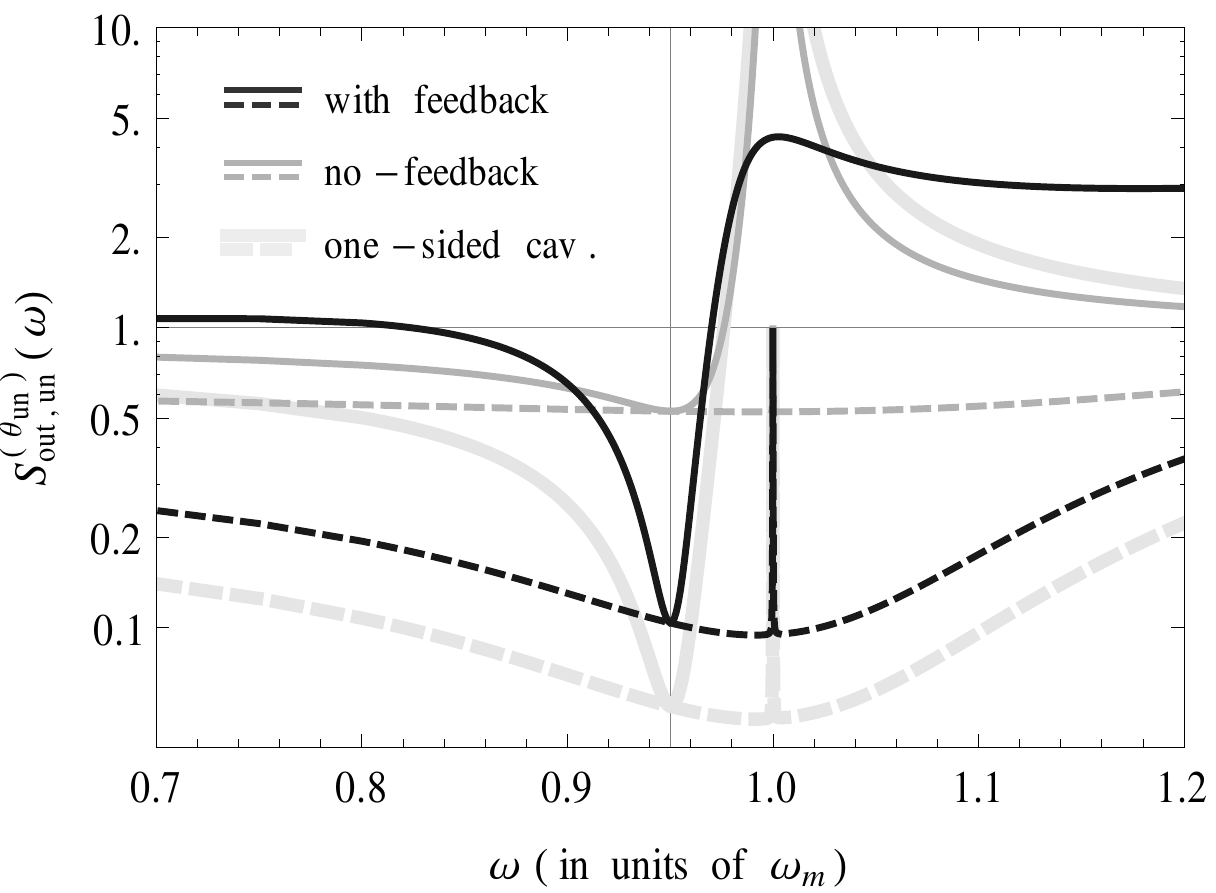}
\caption{Squeezing spectrum $S_{\out,\un}\al{\theta_\un}(\omega)$, of the field reflected by a symmetric cavity ($\kappa_1=\kappa_2$) with the feedback closed in transmission, evaluated for $\omega_{\rm m}=10$\,MHz, $\gamma=10^{-4}\,\omega_{\rm m}$, $T=100$\,mK, $\Delta=0$, $\kappa=\kappa_1+\kappa_2=\omega_{\rm m}$, $\tau_\fb=0.1\,\omega_{\rm m}$, $G=0.5\,\omega_{\rm m}$, $\eta=1$, and the flat feedback filter of Eq.~\rp{gfbflat}.
The dark lines are with feedback, the thin light lines are without feedback, and the thick light lines 
correspond to a single--sided cavity with equal total decay rate and no feedback.
The solid lines are found for the values of $\theta_\un$, $\theta_\fb$, $\bar g_\fb$ and $\phi_\fb$ which optimize the squeezing for the values given above and at the frequency corresponding to the vertical line. The dashed lines are found by optimizing the values of $\theta_\un$, $\theta_\fb$, $\bar g_\fb$ and $\phi_\fb$ for all the frequencies.
}\label{squeezing1}
\end{figure}
\begin{figure}[t!]
\includegraphics[width=8cm]{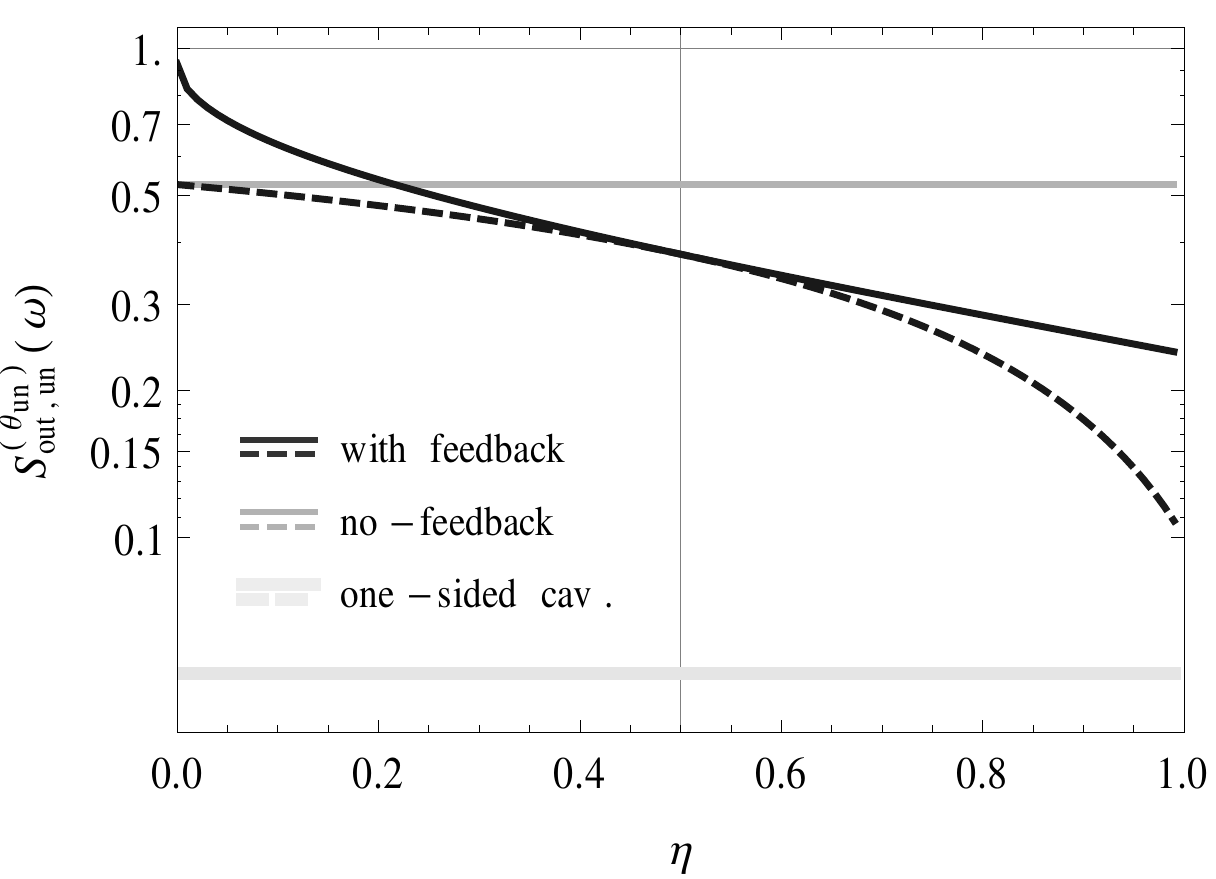}
\caption{Squeezing spectrum $S_{\out,\un}\al{\theta_\un}(\omega)$, of the reflected field with the feedback closed in transmission,  at the frequency indicated by the vertical line in Fig.~\ref{squeezing1}, as a function of the detection efficiency $\eta$.
The solid lines are found for the values of $\theta_\un$, $\theta_\fb$, $\bar g_\fb$ and $\phi_\fb$ which optimize the squeezing at the value of $\eta$ indicated by the vertical line. The dashed lines are found by optimizing the values of $\theta_\un$, $\theta_\fb$, $\bar g_\fb$ and $\phi_\fb$ for all the values of $\eta$. The other parameters and line-styles are as in Fig.~\ref{squeezing1}.
}\label{squeezing2}
\end{figure}
\begin{figure}[t!]
\includegraphics[width=8.5cm]{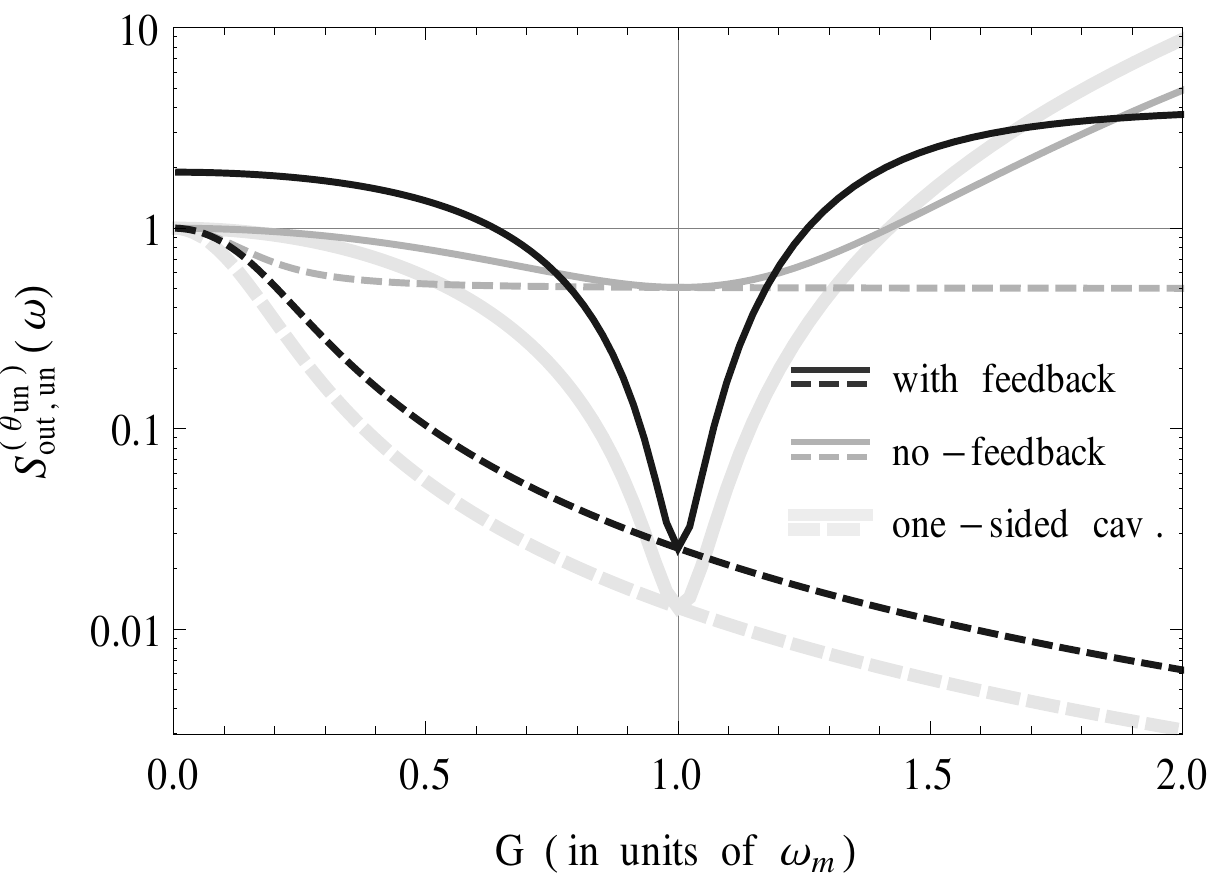}
\caption{Squeezing spectrum $S_{\out,\un}\al{\theta_\un}(\omega)$, of the reflected field with the feedback closed in transmission,  at the frequency indicated by the vertical line in Fig.~\ref{squeezing1}, as a function of the optomechanical coupling strength $G$.
The solid lines are found for the values of $\theta_\un$, $\theta_\fb$, $\bar g_\fb$ and $\phi_\fb$ which optimize the squeezing at the value of $G$ indicated by the vertical line. The dashed lines are found by optimizing the values of $\theta_\un$, $\theta_\fb$, $\bar g_\fb$ and $\phi_\fb$ for all the values of $G$. The other parameters and line-styles are as in Fig.~\ref{squeezing1}.
}\label{squeezing3}
\end{figure}
\begin{figure}[t!]
\includegraphics[width=8.5cm]{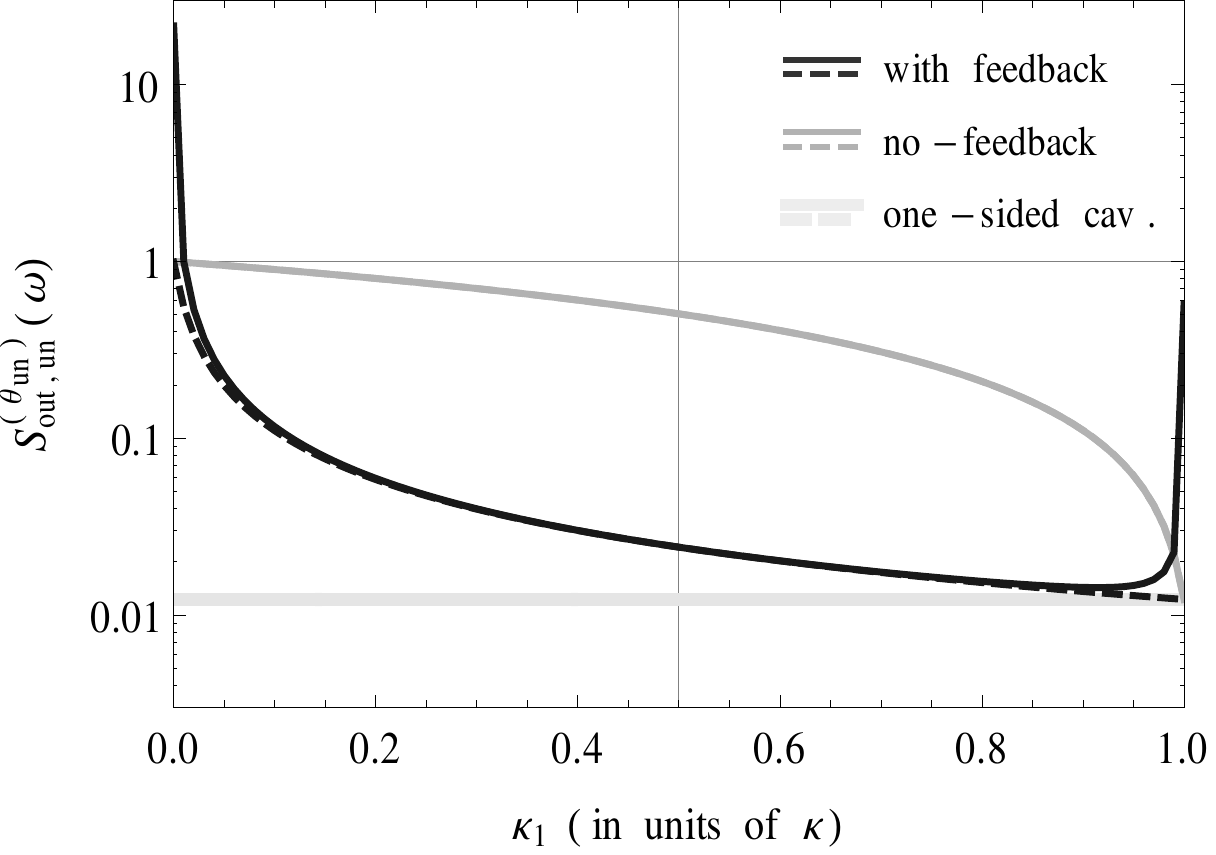}
\caption{Squeezing spectrum $S_{\out,\un}\al{\theta_\un}(\omega)$, of the reflected field with the feedback closed in transmission,  at the frequency indicated by the vertical line in Fig.~\ref{squeezing1}, as a function of the decay rate of the first mirror $\kappa_1$ (with constant $\kappa$).
The solid lines are found for the values of $\theta_\un$, $\theta_\fb$, $\bar g_\fb$ and $\phi_\fb$ which optimize the squeezing at the value of $\kappa_1$ indicated by the vertical line. The dashed lines are found by optimizing the values of $\theta_\un$, $\theta_\fb$, $\bar g_\fb$ and $\phi_\fb$ for all the values of $\kappa_1$. The other parameters and line-styles are as in Fig.~\ref{squeezing1}.
}\label{squeezing4}
\end{figure}

Another fundamental achievement of cavity optomechanics is ponderomotive squeezing, i.e., squeezing of light due to the nonlinear interaction with a mechanical element.
Here we show that feedback--controlled cavities can be exploited to enhance ponderomotive squeezing under certain conditions.

To be specific, we study a cavity--optomechanical system where the light of one cavity output, with decay rate $\kappa_\fb$,  is used to close a feedback loop, and we focus on the properties of the light lost by the cavity from an additional unused cavity output at decay rate $\kappa_\un$ (see Fig.~\ref{schema3}).
We demonstrate that it is possible to achieve stronger squeezing of the unused output light with respect to that achievable with no feedback, but otherwise under the same conditions.
%

Let us consider a quadrature of light of the unused cavity output at phase $\theta_\un$
\begin{eqnarray}
\wt X_{\out,\un}\al{\theta_\un}(\omega)=\sqrt{2\,\kappa_\un}\ \wt X\al{\bar\theta_\un-\phi_c}(\omega)-\wt X_{\inn,\un}\al{\bar\theta_\un}(\omega)\ ,
\end{eqnarray}
with $\bar\theta_\un=\theta_\un+\phi_{\out,\un}$ and $\phi_{\out,\un}$ the phase shift between input and unused output fields.
It can be expressed in terms of
the photocurrent and of the operators without feedback
as
\begin{eqnarray}\label{Xoutf}
\wt X_{\out,\un}\al{\theta_\un}(\omega)
&=&\wt X_{\out,\un}^{\circ\, \pt{\theta_\un}}(\omega)+\frac{1}{\sqrt{\eta}}\ K(\omega)\ \ee^{\ii\,\phi_K(\omega)}
\ \wt i^\circ(\omega)\ ,
\nn\\
\end{eqnarray}
where we have introduced the parameter
\begin{eqnarray}
K(\omega)\ \ee^{\ii\,\phi_K(\omega)}=2\,\sqrt{\eta}\ \zeta_{\om,\un}\al{\theta_\un}(\omega)\ \wt h_\fb(\omega)\ \lambda_{\om,\fb}(\omega)\ ,
\end{eqnarray}
with $K(\omega)$ real and positive, and $\phi_K(\omega)$ real.
Here $\zeta_{\om,\un}\al{\theta_\un}(\omega)$ is the function that describes how input noise fluctuations from the first mirror are transferred to the unused output  such that $\wt X_{\out,\un}\al{\theta_\un}(\omega)=\zeta_{\om,\un}\al{\theta_\un}(\omega)\ \wt X_{\rm in,1}(\omega)+\cdots$, where the dots indicate terms proportional to other input noise operators. Its explicit form is analogous to the one of the transfer function in Eq.~\rp{zetafb_2}, where, now, the cavity response function $\zeta_c\al{\bar\theta_\fb}(\omega)$ includes also the modification due to the optomechanical interaction analogous to that reported in Eq.~\rp{zetaom},
and with the roles of transmission and reflection exchanged.

Using the definition of the photocurrent $\wt i^\circ(\omega)=\sqrt{2\,\kappa_\fb}\ \wt X_{\out,\fb}^{\circ\, \pt{\theta_\fb}}(\omega)+\sqrt{1-\eta}\ \wt X_v(\omega)$ we also find that
Eq.~\rp{Xoutf} can be rewritten as
\begin{eqnarray}
\wt X_{\out,\un}\al{\theta_\un}(\omega)&=&
\sqrt{1+K(\omega)^2}\ \wt X_\out^\circ(\omega)
\nn\\&&\hspace{1cm}
+K(\omega)\,\ee^{\ii\,\phi_K(\omega)}\sqrt{\frac{1-\eta}{\eta}}\ \wt X_v(\omega)\ ,
\end{eqnarray}
 where we have introduced the combined quadrature
\begin{eqnarray}\label{Xtot}
\wt X_\out^\circ(\omega)=\frac{\wt X_{\out,\un}^{\circ\, \pt{\theta_\un}}(\omega)+K(\omega)\,\ee^{\ii\,\phi_K(\omega)}\ \wt X_{\out,\fb}^{\circ\, \pt{\theta_\fb}}(\omega)}{\sqrt{1+K(\omega)^2}}\ ,
\end{eqnarray}
which is a linear combination of the two output quadratures. The ponderomotive squeezing spectrum of a similar combination of quadratures [with $\phi_K(\omega)=0$] has been studied in Ref.~\cite{ZippilliNJP15}, where we have shown that, by properly selecting the coefficients of the linear combination, in a two--sided cavity, the level of squeezing of the combined quadrature reaches the same level of squeezing that can be produced with a single-sided cavity with equal total decay rate. In fact, in a two--sided configuration the cavity light is split and lost through the two output ports, and the two output fields are correlated such that only
their superposition
can reveal the total squeezing that could be produced in a similar system with only one output port.
This fact suggests that by using feedback it is possible to recover part of the light correlations that would otherwise be split between the two outputs.

In particular, by introducing the power spectrum of the photocurrent with no-feedback, $S_i^\circ(\omega)=\eta\ S_{\out,\fb}^{\circ\,(\theta_\fb)}(\omega)+1-\eta$, the power spectrum of the two output quadratures $S_{\out,x}^{\circ\,(\theta_x)}(\omega)\ \delta(\omega+\omega')=\av{\wt X_{\out,x}^{\circ\, \pt{\theta_x}}(\omega)\ \wt X_{\out,x}^{\circ\, \pt{\theta_x}}(\omega')}$, with $x\in\pg{\fb,\un}$, and the spectrum of their correlations with no feedback, defined by the relation $S_{\out,\fb-\un}^{\circ\,(\theta_\fb,\theta_\un)}(\omega)\ \delta(\omega+\omega')=\av{\wt X_{\out,\fb}^{\circ\, \pt{\theta_\fb}}(\omega)\ \wt X_{\out,\un}^{\circ\, \pt{\theta_\un}}(\omega')}$,
we find that the power spectrum of the unused output in the presence of feedback can be expressed as
\begin{eqnarray}\label{Soutf}
S_{\out,\un}\al{\theta_\un}(\omega)&=&S_{\out,\un}^{\circ\,(\theta_\un)}(\omega)+K(\omega)^2\pq{ S_{\out,\fb}^{\circ\,(\theta_\fb)}(\omega)+\frac{1-\eta}{\eta}}
\nn\\&&
+K(\omega)\,\pq{
\ee^{\ii\,\phi_K(\omega)}\ S_{\out,\fb-\un}^{\circ\,(\theta_\fb,\theta_\un)}(\omega)+c.c.
}\ .
\end{eqnarray}
We are interested in identifying the parameter regimes of maximum squeezing, i.e. the parameters for which this expression is minimum.
It turns out that the  minimum of Eq.~\rp{Soutf} is found for the specific phase $\phi_K(\omega)$ for which
\begin{eqnarray}
{\rm Re}\pq{\ee^{\ii\,\phi_K}\ S_{\out,\fb-\un}^{\circ\,(\theta_\fb,\theta_\un)}(\omega)}=-\abs{S_{\out,\fb-\un}^{\circ\,(\theta_\fb,\theta_\un)}(\omega)}\ ,
\end{eqnarray}
so that, for this phase,
\begin{eqnarray}\label{Soutf_2}
S_{\out,\un}\al{\theta_\un}(\omega)&=&S_{\out,\un}^{\circ\,(\theta_\un)}(\omega)+K(\omega)^2\pq{ S_{\out,\fb}^{\circ\,(\theta_\fb)}(\omega)+\frac{1-\eta}{\eta}}
\nn\\&&
-2\ K(\omega)\,\abs{
S_{\out,\fb-\un}^{\circ\,(\theta_\fb,\theta_\un)}(\omega)
}\ .
\end{eqnarray}
The minimum as a function of $K(\omega)$ is instead found for
\begin{eqnarray}
K(\omega)=\frac{\abs{
S_{\out,\fb-\un}^{\circ\,(\theta_\fb,\theta_\un)}(\omega)
}}{
S_{\out,\fb}^{\circ\,(\theta_\fb)}(\omega)+\frac{1-\eta}{\eta}
}\ ,
\end{eqnarray}
with corresponding minimum value
\begin{eqnarray}\label{Soutf_3}
S_{\out,\un}\al{\theta_\un}(\omega)&=&S_{\out,\un}^{\circ\,(\theta_\un)}(\omega)-\frac{\abs{
S_{\out,\fb-\un}^{\circ\,(\theta_\fb,\theta_\un)}(\omega)
}^2
}{
S_{\out,\fb}^{\circ\,(\theta_\fb)}(\omega)+\frac{1-\eta}{\eta}
}\ .
\end{eqnarray}
We further note that the power spectra of the two output quadratures are related by the simple relations
\begin{eqnarray}\label{Soutun}
S_{\out,\fb}^{\circ\,(\theta)}(\omega)=1+\frac{\kappa_\fb}{\kappa_\un}\,\pq{S_{\out,\un}^{\circ\,(\theta)}(\omega)-1}
\end{eqnarray}
and
\begin{eqnarray}\label{Soutfbun}
S_{\out,\fb-\un}^{\circ\,(\theta,\theta')}(\omega)=\sqrt{\frac{\kappa_\fb}{\kappa_\un}}\,\pq{S_{\out,\un}^{\circ\,(\theta,\theta')}(\omega)-\ee^{-\ii(\theta-\theta')}}\ ,
\end{eqnarray}
where $S_{\out,\un}^{\circ\,(\theta,\theta')}(\omega)\ \delta(\omega+\omega')=\av{\wt X_{\out,\un}^{\circ\,(\theta)}(\omega)\ \wt X_{\out,\un}^{\circ\,(\theta')}(\omega')}$.
These relations are direct consequences of the input--output relations defined in Eq.~\rp{aomega0}~\cite{note1}.

In the case in which the feedback phase $\theta_\fb$ is equal to the quadrature phase $\theta_\un$, it is convenient to introduce the parameter  $s_{\out,\un}^\circ(\omega)$ defined by the relation $S_{\out,\un}^{\circ\,(\theta_\un)}(\omega)=1-s_{\out,\un}^\circ(\omega)$ for the spectrum without feedback [such that it is squeezed when $s_{\out,\un}^\circ(\omega)>0$], and, using also Eqs.~\rp{Soutun} and \rp{Soutfbun}, we can rewrite the corresponding optimized squeezing spectrum with feedback defined in Eq.~\rp{Soutf_3}, as
\begin{eqnarray}\label{Soutf_4}
S_{\out,\un}\al{\theta_\un}(\omega)=1-\frac{s_{\out,\un}^\circ(\omega)}{1-\eta\,\frac{\kappa_\fb}{\kappa_\un}\ s_{\out,\un}^\circ(\omega)}\ .
\end{eqnarray}
We note that the parameter  $s_{\out,\un}^\circ(\omega)$ fulfils the relation $s_{\out,\un}^\circ(\omega)<\kappa_\un/\kappa$, where $\kappa$ is the total cavity decay rate.
This is due to the fact that only a fraction $\sqrt{\kappa_\un/\kappa}$ of the cavity field leaks through the output with decay rate $\kappa_\un$, so that,
when compared with the corresponding power spectrum for a single--sided cavity, with equal total decay rate, $S_{\rm sing}^{\circ\,(\theta_\un)}(\omega)=1-s_{\rm sing}^\circ(\omega)$, where $s_{\rm sing}^\circ(\omega)<1$, one finds $s_{\out,\un}^\circ(\omega)=s_{\rm sing}^\circ(\omega)\, \kappa_\un/\kappa$~\cite{note1}.
%
%
%
Furthermore, this result implies that, if the quadrature without feedback is squeezed, i.e. $s_{\out,\un}^\circ(\omega)>0$, then the relation
$S_{\out,\un}\al{\theta_\un}(\omega)<S_{\out,\un}^{\circ\,(\theta_\un)}(\omega)$ is always true, meaning that the squeezing can \emph{always} be increased by using feedback.

However, we observe that the value of $S_{\out,\un}\al{\theta_\un}(\omega)$ in a two-sided cavity is always larger than the corresponding squeezing spectrum achievable without feedback in a system with a single dissipation channel, but same total dissipation rate and otherwise equivalent. In this case, in fact, as stated above,
$s_{\rm sing}^\circ(\omega)=s_{\out,\un}^\circ(\omega){\kappa}/{\kappa_\un}$, so that from Eq.~\rp{Soutf_4} we find $S_{\out,\un}\al{\theta_\un}(\omega)\geq S_{\rm sing}^{\circ\,(\theta_\un)}(\omega)$, with the equal sign achieved when $\eta=1$, $\kappa=\kappa_\fb+\kappa_\un$ (no additional dissipation) and $s_{\rm sing}^\circ(\omega)=1$ such that $S_{\out,\un}\al{\theta_\un}(\omega)=0$.

These results are described by Figs.~\ref{squeezing1}--\ref{squeezing4} where we plot the squeezing spectrum of the unused output field with (dark lines) and without (thin light lines) feedback. The thick light lines are the results of the single-sided cavity and no feedback. The solid lines are evaluated by optimizing the feedback parameters at the specific value of the x-axis identified by the vertical lines in each plot, while the dashed lines are optimized at every point. The results show that the feedback can reduce the power spectrum (increase the squeezing), and that it is lower--bounded by the result of the single-sided cavity. In particular, Fig.~\ref{squeezing4} shows
how the improvement due to feedback disappears progressively as the ratio $\kappa_1/\kappa_2$ increases.
This means that this kind of feedback can not improve the optimal ponderomotive squeezing achievable in a single sided cavity, however it could be useful in realistic situations in which the optical cavity has additional decay channels.

We finally remark that this strategy shares similarities with related protocols based on coherent feedback~\cite{Gough2009,Iida2012,Kraft2016,Nemet2016}, and
it is not solely useful for optomechanical systems. In fact, feedback--controlled light can also be applied to, for example, an optical parametric oscillator in a two--sided cavity and achieve a similar improvement of the resulting squeezing.

\section{Conclusions and Outlook}\label{conclusion}

The results presented in this work demonstrate that feedback--controlled light may play a significant role as a novel efficient tool to manipulate cavity--optomechanical systems (and possibly other quantum systems
\cite{Wagner}).

We have described how to design the fluctuations of squashed and anti-squashed light in order to effectively reduce the cavity linewidth and to observe coherent optomechanical oscillations in weakly coupled systems;
to control interference effects which allow for enhanced optomechanical sideband cooling;
and to increase the ponderomotive squeezing that can be extracted by an optomechanical system with a two-sided cavity.

The flexibility and the simplicity of application make this approach particularly appealing.
However, the full potentiality of this technique needs to be further explored.
A prominent question is whether this approach can be adapted to the preparation of other quantum states of mechanical resonators. A specific example is the preparation of two--mode squeezing (entanglement) of two mechanical modes~\cite{Ockeloen-Korppi,Woolley,Li15} with multi-frequency driving fields~\cite{Mari}. In these cases larger and more robust entanglement is observed at smaller cavity decay rates. In-loop cavities could therefore be useful in a way similar to what has already been discussed with coherent feedback~\cite{Li17}.
More generally, it seems interesting to explore the consequences of the fact that
in-loop fields permit to promote an optomechanical system to the strong coupling and
to the resolved sideband regime even if the cavity linewidth is naturally large.
Many theoretical proposals that operate in these regimes
could benefit from in-loop cavities. An intriguing example is the implementation of quantum heat engines which make use of polariton excitations in an optomechanical system~\cite{Zhang}.
In this case the realization of the heat engine can be eased by feedback and, at the same time, the additional correlated feedback noise could possibly be exploited to achieve enhanced efficiency, as demonstrated in similar systems with correlated baths~\cite{Klaers}.
Another interesting example is the study of phonon-based topological dynamics similar to what has been discussed in
\cite{Xu} where feedback-controlled light may allow the realization of similar processes even with resonators which are not naturally in the resolved sideband regime.

\

\acknowledgments
We acknowledge the support of the European Union's
Horizon 2020 research and innovation program under grant
agreement No. 732894 (FET Proactive HOT).






\begin{thebibliography}{99}

\bibitem{Bowen}
W. P. Bowen, G. J. Milburn, {\it Quantum Optomechanics}, CRC Press, Taylor \& Francis Group (2016).

\bibitem{Aspelmeyer}
M. Aspelmeyer, T. J. Kippenberg, and F. Marquardt, {\it Cavity optomechanics}, Rev. Mod. Phys. {\bf 86}, 1391 (2014).

\bibitem{Metcalfe} M. Metcalfe, {\it Applications of cavity optomechanics}, Applied Physics Reviews {\bf 1}, 031105 (2014).




\bibitem{Fiore2011} V. Fiore, Y. Yang, M. C. Kuzyk, R. Barbour, L. Tian, and H. Wang, {\it Storing Optical Information as a Mechanical Excitation in a Silica Optomechanical Resonator}, Phys. Rev. Lett. {\bf 107}, 133601 (2011).

\bibitem{Marquardt2006} F. Marquardt, J. G. E. Harris, and S. M. Girvin, {\it Dynamical Multistability Induced by Radiation Pressure in High-Finesse Micromechanical Optical Cavities}, Phys. Rev. Lett. {\bf 96}, 103901 (2006).

\bibitem{Stannigel} K. Stannigel, P. Komar, S. J. M. Habraken, S. D. Bennett, M. D. Lukin, P. Zoller, and P. Rabl, {\it Optomechanical Quantum Information Processing with Photons and Phonons}, Phys. Rev. Lett. {\bf 109}, 013603 (2012).

%
%
%
%


\bibitem{Chen2013} Y. Chen, {\it Macroscopic quantum mechanics: theory and experimental concepts of optomechanics}, J. Phys. B: At. Mol. Opt. Phys. {\bf 46}, 104001 (2013).

\bibitem{Serafini2012} A. Serafini, {\it Feedback Control in Quantum Optics: An Overview of Experimental Breakthroughs and Areas of Application}, ISRN Optics {\bf 2012}, 1 (2012).


\bibitem{Zhang2017} J. Zhang, Y. Liu, R.-B. Wu, K. Jacobs, and F. Nori, {\it Quantum feedback: Theory, experiments, and applications}, Phys. Rep. {\bf 679}, 1 (2017).


\bibitem{Wilson2015} D. J. Wilson, V. Sudhir, N. Piro, R. Schilling, A. Ghadimi, and T. J. Kippenberg, {\it Measurement-based control of a mechanical oscillator at its thermal decoherence rate}, Nature {\bf 524}, 325 (2015).

\bibitem{Sudhir} V. Sudhir, {\it Quantum limits on measurement and control of a mechanical oscillator} (Springer Berlin Heidelberg, New York, NY, 2017).

\bibitem{Schafermeier} C. Sch\"afermeier, H. Kerdoncuff, U. B. Hoff, H. Fu, A. Huck, J. Bilek, G. I. Harris, W. P. Bowen, T. Gehring, and U. L. Andersen, {\it Quantum enhanced feedback cooling of a mechanical oscillator using nonclassical light}, Nature Communications {\bf 7}, 13628 (2016).

\bibitem{Rossi2018} M. Rossi, D. Mason, J. Chen, Y. Tsaturyan, and A. Schliesser, {\it Measurement-based quantum control of mechanical motion}, arXiv:1805.05087 [quant-ph] (2018).

\bibitem{Zhang2009} J. Zhang, Y. Liu, and F. Nori, {\it Cooling and squeezing the fluctuations of a nanomechanical beam by indirect quantum feedback control}, Phys. Rev. A{\bf 79}, 052102 (2009).

\bibitem{Shapiro87} J. H. Shapiro, P. Kumar, B. E. A. Saleh, M. C. Teich, G. Saplakoglu, and S.-T. Ho, {\it Theory of light detection in the presence of feedback}, JOSA B {\bf 4}, 1604 (1987).


\bibitem{Yamamoto86} Y. Yamamoto, N. Imoto, and S. Machida, {\it Amplitude squeezing in a semiconductor laser using quantum nondemolition measurement and negative feedback}, Phys. Rev. A {\bf 33}, 3243 (1986).



\bibitem{Wiseman98} H. M. Wiseman, {\it In-Loop Squeezing Is Like Real Squeezing to an In-Loop Atom}, Phys. Rev. Lett. {\bf 81}, 3840 (1998).

\bibitem{Wiseman99} H. M. Wiseman, {\it Squashed states of light: theory and applications to quantum spectroscopy}, Journal of Optics B: Quantum and Semiclassical Optics 1, 459 (1999).



\bibitem{Rossi} M. Rossi, N. Kralj, S. Zippilli, R. Natali, A. Borrielli, G. Pandraud, E. Serra, G. Di Giuseppe, and D. Vitali, {\it Enhancing Sideband Cooling by Feedback-Controlled Light}, Phys. Rev. Lett. {\bf 119}, 123603 (2017).

\bibitem{Kralj} N. Kralj, M. Rossi, S. Zippilli, R. Natali, A. Borrielli, Gregory Pandraud, E. Serra, G. D. Giuseppe, and D. Vitali, {\it Enhancement of three-mode optomechanical interaction by feedback-controlled light}, Quantum Sci. Technol. {\bf 2}, 034014 (2017).

\bibitem{Rossi2} M. Rossi, N. Kralj, S. Zippilli, R. Natali, A. Borrielli, G. Pandraud, E. Serra, G. Di Giuseppe, and D. Vitali, {\it Normal-Mode Splitting in a Weakly Coupled Optomechanical System}, Phys. Rev. Lett. {\bf 120}, 073601 (2018).

\bibitem{ZippilliNJP15}
S. Zippilli, G. Di Giuseppe, and D. Vitali, {\it Entanglement and squeezing of continuous-wave stationary light}, New J. Phys.{\bf 17}, 043025 (2015).

\bibitem{electronicNoise}%
The effect of the electronic noise of the detection apparatus can be included in terms of a zero mean, white noise stochastic term [i.e. $\av{F_e(t)}=0$ and $\av{F_e(t) F_e(t')}\sim S_e\,\delta(t-t')$] in the photocurrent, such that, in the case of direct photodetection,
$I_d(t)=A_d\da(t)\,A_d(t)+F_e(t)$, where $A_d(t)=\sqrt{\eta_d}\,A_\inn(t)+\sqrt{1-\eta_d}\,v_d(t)$ is the operator for the detected field which includes the detection efficiency $\eta_d$ and the corresponding additional noise operator $v_d$, which describes vacuum noise. 
Retaining only linear terms in the fluctuations, the photocurrent fluctuations are described by $i_d(t)\simeq \alpha_\inn\,\eta_d\,X_\inn(t)+\alpha_\inn\,\sqrt{\eta_d(1-\eta_d)}\, X_{v_d}(t)+F_e(t)$. 
If the detected field is coherent, the corresponding power spectrum is the sum of electronic plus shot noise $S_{i_d}=S_e+S_{sn}$, with $S_{sn}=\alpha_\inn^2\ \eta_d$. By normalizing the photocurrent such that the corresponding power spectrum is equal to one, i.e  $i(t)=i_d(t)/\sqrt{S_e+S_{sn}}$, we find the expression reported in Eq.~\rp{i} (with $\theta_\fb=0$), where $X_v(t)=\pq{\sqrt{S_{sn}(1-\eta_d)}\,X_{v_d}(t)+F_e(t)}/\sqrt{S_{sn}(1-\eta_d)+S_e}$. Similar considerations can be easily generalized to the case of homodyne detection.
%

\bibitem{Sheard} B. S. Sheard, M. B. Gray, B. J. J. Slagmolen, J. H. Chow, and D. E. McClelland, {\it Experimental demonstration of in-loop intracavity intensity-noise suppression}, IEEE Journal of Quantum Electronics {\bf 41}, 434 (2005).



%
%
%
\bibitem{GardinerZoller}
C. W. Gardiner, P. Zoller, {\it Quantum Noise}, Heidelberg Springer (2004).

\bibitem{Genes08} C. Genes, D. Vitali, P. Tombesi, S. Gigan, and M. Aspelmeyer, {\it Ground-state cooling of a micromechanical oscillator: Comparing cold damping and cavity-assisted cooling schemes}, Phys. Rev. A \textbf{77}, 033804 (2008).





\bibitem{note1}
In the case of a cavity with various output channels (and no feedback), a quadrature of the field at the specific output $j$ with decay rate $\kappa_j$ fulfils the relation $X_{\out,j}=\sqrt{2\,\kappa_j}\ X-X_{\inn,j}$ where $X$ is a cavity quadrature and $X_{\inn,j}$ the corresponding input noise quadrature.
Hence, the corresponding power spectrum takes the form $S_{X_{\out,j},X_{\out,j}}=2\,\kappa_j\ S_{X,X}+1-2\sqrt{2\,\kappa_j}{\rm Re}\pq{S_{X,X_{\inn,j}}}$, where we assume vacuum input noise so that $S_{X_{\inn,j},X_{\inn,j}}=1$. Moreover since $X=\sqrt{2\,\kappa_j}\ \chi_c\ X_{\inn,j}+\cdots$, where the dots stand for terms proportional to other input noise operators, and $\chi_c$ is the cavity susceptibility, we can introduce $\bar X=\chi_c\ X_{\inn,j}+\cdots$ and the corresponding cross--power spectrum $S_{\bar X,X_{\inn}}$ (which is the same for all the input noise operators, hence we can drop the index $j$) so that $S_{X,X_{\inn,j}}=\sqrt{2\,\kappa_j}\ S_{\bar X,X_\inn}$, and eventually
$S_{X_{\out,j},X_{\out,j}}=1+2\,\kappa_j\pg{ S_{X,X}-2{\rm Re}\pq{S_{\bar X,X_\inn}}}$. The power spectrum of different outputs are distinguished only by the specific value of the corresponding decay rate $\kappa_j$ in the previous expression. In the case of a single-sided cavity the power spectrum of the cavity output is given by the previous expression with the total decay rate $\kappa$ in place of the $\kappa_j$.



\bibitem{Gough2009} J. E. Gough and S. Wildfeuer, {\it Enhancement of field squeezing using coherent feedback}, Phys. Rev. A {\bf 80}, 042107 (2009).

\bibitem{Iida2012} S. Iida, M. Yukawa, H. Yonezawa, N. Yamamoto, and A. Furusawa, {\it Experimental Demonstration of Coherent Feedback Control on Optical Field Squeezing}, IEEE Transactions on Automatic Control {\bf 57}, 2045 (2012).

%
%

\bibitem{Kraft2016} M. Kraft, S. M. Hein, J. Lehnert, E. Schöll, S. Hughes, and A. Knorr, {\it Time-delayed quantum coherent Pyragas feedback control of photon squeezing in a degenerate parametric oscillator}, Phys. Rev. A {\bf 94}, 023806 (2016).

\bibitem{Nemet2016} N. N\'emet and S. Parkins, {\it Enhanced optical squeezing from a degenerate parametric amplifier via time-delayed coherent feedback}, Phys. Rev. A {\bf 94}, 023809 (2016).




\bibitem{Wagner}
T. Wagner, P. Strasberg, J. C. Bayer, E. P. Rugeramigabo, T. Brandes, and R. J. Haug, {\it Strong suppression of shot noise in a feedback-controlled single-electron transistor}, Nature Nanotechnology {\bf 12}, 218 (2016).



\bibitem{Ockeloen-Korppi}
C. F. Ockeloen-Korppi, E. Damsk\"agg, J.-M. Pirkkalainen, M. Asjad, A. A. Clerk, F. Massel, M. J. Woolley, and M. A. Sillanp\"a\"a, {\it Stabilized entanglement of massive mechanical oscillators}, Nature {\bf 556}, 478 (2018).


\bibitem{Woolley}
M. J. Woolley and A. A. Clerk, {Two-mode squeezed states in cavity optomechanics via engineering of a single reservoir}, Phys. Rev. A {\bf 89}, 063805 (2014)

\bibitem{Li15}
J. Li, I. M. Haghighi, N. Malossi, S. Zippilli, and D. Vitali, {Generation and detection of large and robust entanglement between two different mechanical resonators in cavity optomechanics}, New J. Phys. {\bf 17}, 103037 (2015)

\bibitem{Mari} A. Mari and J. Eisert, {Gently Modulating Optomechanical Systems}, Phys. Rev. Lett. {\bf 103}, 213603 (2009).

\bibitem{Li17}
J. Li, G. Li, S. Zippilli, D. Vitali, and T. Zhang, {Enhanced entanglement of two different mechanical resonators via coherent feedback}, Phys. Rev. A {\bf 95}, 043819 (2017)

\bibitem{Zhang}
K. Zhang, F. Bariani, and P. Meystre, {Quantum Optomechanical Heat Engine}, Phys. Rev. Lett. {\bf 112}, 150602 (2014)

\bibitem{Klaers}
J. Klaers, S. Faelt, A. Imamoglu, and E. Togan, {Squeezed Thermal Reservoirs as a Resource for a Nanomechanical Engine beyond the Carnot Limit}, Phys. Rev. X {\bf 7}, 031044 (2017)

\bibitem{Xu}
H. Xu, D. Mason, L. Jiang, and J. G. E. Harris, {Topological energy transfer in an optomechanical system with exceptional points}, Nature {\bf 537}, 80 (2016)


%
%
%
%
%


%
%
%
%
%
%
%
%
%
%
%
%
%
%
%
%
%
%
%
%
%
%
%
%
%
%
%
%
%
%
%
%
%
%
%
%
%
%
%
%
%
%
%
%
%
%
%
%
%
%
%
%
%

\end{thebibliography}
\end{document}